\documentclass{article}
\pdfoutput=1

\usepackage{mathtools}
\usepackage{booktabs}
\usepackage[english]{babel}
\usepackage{amsmath,amssymb,amsbsy,amstext, amsthm, simplewick, amsfonts}
\usepackage{graphicx}
\usepackage[small]{caption}
\usepackage{siunitx}
\usepackage{upgreek}
\usepackage{framed}
\usepackage{wrapfig}
\usepackage{multirow}
\usepackage{bbm}
\usepackage[numbers,sort&compress]{natbib}
\usepackage[svgnames,dvipsnames,x11names]{xcolor}
\usepackage[utf8x]{inputenc}
\usepackage{selinput}
\usepackage{bm}
\usepackage{float}
\usepackage{geometry}
\usepackage{yfonts}
\usepackage{caption}
\usepackage{subcaption}
\usepackage{sidecap}
\usepackage{longtable}
\usepackage{anyfontsize}

\usepackage{epstopdf}
\usepackage{cancel}
\usepackage{tcolorbox}

\def\xyma{\xymatrix@M.7em}
\def\xymas{\xymatrix@M.1em}

\newcommand{\Comment}[1]{{}}
\definecolor{darkblue}{rgb}{0.15,0.35,0.55}
\definecolor{reddish}{rgb}{0.65, 0.2, 0.2}
\definecolor{darkgreen}{RGB}{50,150,0}
\definecolor{greyish2}{rgb}{.96,.96,.96}
\usepackage[linktocpage=true]{hyperref}
\hypersetup{
colorlinks=true,
citecolor=darkblue,
linkcolor=reddish,
urlcolor=darkblue,
pdfauthor={},
pdftitle={},
pdfsubject={}
}

\usepackage{cleveref}

\usepackage{pgfornament}

\DeclareFontFamily{OT1}{rsfs10}{}
\DeclareFontShape{OT1}{rsfs10}{m}{n}{ <-> rsfs10 }{}
\DeclareMathAlphabet{\mathscript}{OT1}{rsfs10}{m}{n}

\def\gsim{ \lower .75ex \hbox{$\sim$} \llap{\raise .27ex \hbox{$>$}} }
\def\lsim{ \lower .75ex \hbox{$\sim$} \llap{\raise .27ex \hbox{$<$}} }
\def\be{\begin{equation}}
\def\ee{\end{equation}}
\def\bea{\begin{equation}}
\def\eea{\end{equation}}

\newcommand{\dd}{\mathrm{d}}

\newcommand{\mn}{{\mu\nu}}
\newcommand{\ab}{{\alpha\beta}}

\newcommand{\Ij}{{ij}}
\newcommand{\rs}{r_\mathrm{s}}

\newcommand{\sch}{\mathrm{Sch}}

\newcommand{\Mp}{M_\mathrm{Pl}}

\usepackage{latexsym,amsmath,amssymb,epsfig}

\usepackage{tikz}
\usetikzlibrary{decorations}
\pgfdeclaredecoration{complete sines}{initial}
{
    \state{initial}[
        width=+0pt,
        next state=upsine,
        persistent precomputation={\pgfmathsetmacro\matchinglength{
            \pgfdecoratedinputsegmentlength / int(\pgfdecoratedinputsegmentlength/\pgfdecorationsegmentlength)}
            \setlength{\pgfdecorationsegmentlength}{\matchinglength pt}
        }] {}
    \state{upsine}[width=\pgfdecorationsegmentlength,next state=downsine]{
        \pgfpathsine{\pgfpoint{0.25\pgfdecorationsegmentlength}{0.5\pgfdecorationsegmentamplitude}}
        \pgfpathcosine{\pgfpoint{0.25\pgfdecorationsegmentlength}{-0.5\pgfdecorationsegmentamplitude}}
    }
    \state{downsine}[width=\pgfdecorationsegmentlength,next state=upsine]{
        \pgfpathsine{\pgfpoint{0.25\pgfdecorationsegmentlength}{-0.5\pgfdecorationsegmentamplitude}}
        \pgfpathcosine{\pgfpoint{0.25\pgfdecorationsegmentlength}{0.5\pgfdecorationsegmentamplitude}}
}
    \state{final}{}
}

\definecolor{greyish}{rgb}{.90,.90,.90}
\definecolor{greyish2}{rgb}{.96,.96,.96}
\usepackage{xcolor,colortbl}
\usepackage{tcolorbox}

\usepackage[all]{xy}

\usepackage{ytableau}

\numberwithin{equation}{section}

\begin{document}
\renewcommand{\thefootnote}{\fnsymbol{footnote}}
\vspace{0truecm}
\thispagestyle{empty}

\vspace*{-0.3cm}

\begin{center}
{\fontsize{21}{18} \bf Symmetries of Vanishing Nonlinear}\\[14pt]
{\fontsize{21}{18} \bf   Love Numbers of Schwarzschild Black Holes}
\end{center}

\vspace{.15truecm}

\begin{center}
{\fontsize{13}{18}\selectfont
Oscar Combaluzier--Szteinsznaider,${}^{\rm a}$\footnote{\texttt{combaluzier-szteinsznaider@apc.in2p3.fr}} 
Lam Hui,${}^{\rm b}$\footnote{\texttt{lh399@columbia.edu}} 
Luca Santoni,${}^{\rm a}$\footnote{\texttt{santoni@apc.in2p3.fr}} 
\\
  \vspace{.3cm}
Adam R. Solomon,${}^{\rm c,d}$\footnote{\texttt{adam.solomon@gmail.com}}
and 
Sam S. C. Wong${}^{\rm e}$\footnote{\texttt{samwong@cityu.edu.hk}}
}
\end{center}
\vspace{.4truecm}

\centerline{{\it ${}^{\rm a}$Universit\'e Paris Cit\'e, CNRS, Astroparticule et Cosmologie,}}
 \centerline{{\it 10 Rue Alice Domon et L\'eonie Duquet, F-75013 Paris, France}}
 
  \vspace{.3cm}

 \centerline{{\it ${}^{\rm b}$Center for Theoretical Physics, Department of Physics, Columbia University,}}
 \centerline{{\it 538 West 120th Street, New York, NY 10027, U.S.A.}} 
 
  \vspace{.3cm}

\centerline{{\it ${}^{\rm c}$Department of Physics and Astronomy, McMaster University,}}
 \centerline{{\it 1280 Main Street West, Hamilton ON, Canada}} 
 
  \vspace{.3cm}

\centerline{{\it ${}^{\rm d}$Perimeter Institute for Theoretical Physics,}}
\centerline{{\it 31 Caroline Street North, Waterloo ON, Canada}}

 \vspace{.3cm}

\centerline{{\it ${}^{\rm e}$Department of Physics, City University of Hong Kong,}}
\centerline{{\it Tat Chee Avenue, Kowloon, Hong Kong SAR, China}}
 \vspace{.25cm}

\vspace{.3cm}
\begin{abstract}

\noindent
The tidal Love numbers parametrize the conservative induced tidal
response of self-gravitating objects. It is well established that
asymptotically-flat black holes in four-dimensional general relativity
have vanishing Love numbers. In linear perturbation theory, this
result was shown to be a consequence of ladder symmetries acting on
black hole perturbations. In this work, we show that a black hole's
tidal response induced by a static, parity-even tidal field vanishes
for all multipoles to all orders in perturbation theory. Our strategy
is to focus on static and axisymmetric spacetimes for which the
dimensional reduction to the fully nonlinear Weyl solution is
well-known.
{We define the nonlinear Love numbers using the
point-particle effective field theory, matching with the Weyl solution
to show that an infinite subset of the static, parity-even Love number
couplings vanish, to all orders in perturbation theory.
This conclusion holds even if the tidal field deviates from
axisymmetry.}
Lastly, we discuss the symmetries underlying the vanishing of the nonlinear Love numbers. An $\mathfrak{sl}(2,\mathbb R)$ algebra acting on a covariantly-defined potential furnishes ladder symmetries analogous to those in linear theory. This is because the dynamics of the potential are isomorphic to those of a static, massless scalar on a Schwarzschild background. We comment on the connection between the ladder symmetries and the Geroch group that is well-known to arise from dimensional reduction.

\end{abstract}

\newpage

\setcounter{tocdepth}{2}
\tableofcontents
\newpage
\renewcommand*{\thefootnote}{\arabic{footnote}}
\setcounter{footnote}{0}

\section{Introduction}

The black holes of general relativity famously display an astonishing beauty and simplicity~\cite{Chandrasekhar:1985kt}. Not only do black hole solutions appear to be simple, uniquely characterized in terms of a few external macroscopic parameters (mass, spin, and charge) and constrained by general no-hair theorems~\cite{Israel:1967wq,Carter:1968rr,Carter:1971zc,Wald:1971iw,Hartle:1971qq,Bekenstein:1971hc,Fackerell:1972hg,Price:1972pw,Bekenstein:1995un,Hui:2012qt}, but this simplicity is in turn inherited by their perturbations. One notable example is given by a black hole's linear, static tidal response~\cite{Fang:2005qq,Damour:2009vw,Binnington:2009bb}.
The tidal deformation of  a  compact object in a gravitational theory
is parametrized  in terms of a set of coefficients, which can be distinguished into two classes depending on whether they describe  the  conservative  or the dissipative part of the response, induced by an external long-wavelength gravitational tidal field~\cite{Goldberger:2004jt,Goldberger:2005cd}.
The conservative coefficients are often referred to as tidal Love
numbers, and together with the dissipative numbers they carry relevant
information about the physics of the compact object in question.

It is by now well-known that, in contrast to generic self-gravitating
bodies, the Love numbers of (asymptotically flat) black holes in
four-dimensional general relativity vanish
identically~\cite{Fang:2005qq,Damour:2009vw,Binnington:2009bb,Kol:2011vg,Gurlebeck:2015xpa,Hui:2020xxx,
  Hui:2021vcv,Hui:2022vbh,Rai:2024lho,LeTiec:2020spy,LeTiec:2020bos,Charalambous:2021mea,Rodriguez:2023xjd}.
The mysterious nature of this vanishing is accentuated in the worldline effective field theory (EFT) approach to tidal deformations, where it
translates into the absence of a set of Wilson couplings of quadratic, higher-derivative, static operators 
in the EFT~\cite{Goldberger:2004jt,Goldberger:2006bd,Rothstein:2014sra,Porto:2016pyg,Levi:2018nxp,Goldberger:2022ebt,Goldberger:2022rqf}. This effectively makes black holes indistinguishable (in the static limit) from elementary point particles when seen from long distances, at least as far as linear perturbation theory is concerned. 
This property, which had for a long time been known as an outstanding
naturalness puzzle in the infrared description of compact sources in
gravity~\cite{Rothstein:2014sra,Porto:2016zng}, has recently found an
explanation in terms  of a hidden structure of exact ladder symmetries for
static perturbations around black holes \cite{Hui:2021vcv,Hui:2022vbh,Rai:2024lho,BenAchour:2022uqo,Berens:2022ebl}.\footnote{See
  \cite{Charalambous:2021kcz,Charalambous:2022rre} for a different
  proposal based on symmetries in a near-zone approximation, \cite{Solomon:2023ltn} for the role of electric--magnetic duality in the vanishing of gravitational Love numbers, and \cite{Sharma:2024hlz} for a study of more general spherically symmetric and static backgrounds.
  }
The ladder symmetries constrain  the solution of the linearized perturbations to take the form of simple polynomials in the radial coordinate, and enforce the vanishing of the Love numbers. Their existence is a manifestation of the aforementioned special and elementary nature of black holes in general relativity.

Intriguingly, this is not yet the end of the story. The relation
between vanishing Love numbers and hidden symmetries of general
relativity has to date been established only within the scope of
linear perturbation theory.  However, recent results have shown that
Love numbers of Schwarzschild black holes are zero beyond just linear
theory~\cite{Gurlebeck:2015xpa,Poisson:2020vap,Poisson:2021yau,Riva:2023rcm,Iteanu:2024dvx}
(see also \cite{DeLuca:2023mio} for a scalar field example). Recently
the nonlinear Einstein equations were solved in the static  limit at
quadratic order in the fields  and to all orders in the multipolar
expansion, including both even and odd perturbations
\cite{Iteanu:2024dvx}. The quadratic solutions can in general be
written analytically in closed form as simple, finite polynomials, and
the quadratic Love number couplings vanish at all orders in
derivatives in the worldline EFT, precisely as occurs for linear
perturbations \cite{Iteanu:2024dvx}. These results hint at a putative resummation to an
underlying hidden symmetry at the fully nonlinear level.\footnote{We
  emphasize that we are working in the exact time-independent
  limit. There is also evidence for symmetries at low frequency in the
  near-zone approximation \cite{Hui:2022vbh,Charalambous:2021kcz,Charalambous:2022rre,Perry:2023wmm}.}

A notable example of such a resummation with obvious relevance to the
low-frequency physics of black holes is the Weyl class of
solutions~\cite{Griffiths:2009dfa}. In particular, any
 static,\footnote{Here we mean static as opposed to stationary; both possess a timelike Killing vector, but static spacetimes additionally lack time-space cross terms in the metric \cite{Townsend:1997ku}.}  axially-symmetric vacuum solution in
general relativity may be written as a \emph{Weyl
  metric}~\cite{Weyl:1917gp,Stephani:2003tm},
\begin{equation}\label{eq:weyl}
\dd s^2_4 = -\mathrm{e}^{-\psi}\dd t^2 + \mathrm{e}^\psi\left[\mathrm{e}^{2\gamma}(\dd\rho^2+\dd z^2)+\rho^2\dd\phi^2\right].
\end{equation}
Here $t$ and $\phi$ are coordinates adapted to the temporal and angular Killing vectors, while $\rho$ and $z$ are  the so-called Weyl canonical coordinates.\footnote{The coordinates $(\rho,z)$ can be thought of as cylindrical coordinates in an auxiliary flat space, although we stress that they do not necessarily have such an interpretation in the physical spacetime.}
The potential $\psi$ and conformal factor $\gamma$ are functions of $\rho$ and $z$.
The Weyl formulation has the remarkable property that a solution to the full, nonlinear Einstein equations may be obtained by solving a \emph{linear} equation for $\psi$,\footnote{The conformal factor $\gamma$ is determined by a constraint equation.}
\begin{equation}\label{eq:psi-eom}
\left(\partial_\rho^2+\frac1\rho\partial_\rho+\partial_z^2\right)\psi=0.
\end{equation}

The ansatz \eqref{eq:weyl} has been used in~\cite{Gurlebeck:2015xpa}
to show that all (Weyl) multiple-moments of a tidally deformed
Schwarzschild black hole must match those of an undeformed one,
at the fully nonlinear level for axisymmetric configurations (see also the recent \cite{Barcelo:2024ioe}).
In~\cite{Gurlebeck:2015xpa}, the induced nonlinear deformation of the
black hole was defined in terms of source integrals~\cite{Gurlebeck:2013eia}, following the
Geroch--Hansen definition of asymptotic multipole
moments~\cite{Geroch:1970cc,Geroch:1970cd,Hansen:1974zz,Mayerson:2022ekj}.
This result suggests a large class of nonlinear Love numbers, suitably
defined, are zero. A different strategy, making use of a charged black
hole and a charged particle and taking the vanishing charge limit in
the end, was adopted by \cite{Poisson:2021yau} to deduce the nonlinear
tidal response of black holes.

In this work, we make progress in two main directions. First,
we define the nonlinear Love numbers as
in~\cite{Bern:2020uwk,Riva:2023rcm,Iteanu:2024dvx}  at the level of
the point-particle EFT. This will allow us to systematically define
the nonlinear response of the object in a way that is not affected by
ambiguities, due to e.g.,~coordinate choice or nonlinear mixing. By
performing explicitly the matching with the EFT we will conclude that {the
 nonlinear Love number couplings of an infinite subset of operators involving parity-even
fields only} vanish to all orders in perturbation theory.
An advantage of the EFT matching procedure is that it makes clear,
even though the nonlinear solution used for matching
is axisymmetric, the resulting conclusion of vanishing Love
number couplings goes beyond axisymmetry.
In addition, we show that the ladder symmetries of~\cite{Hui:2021vcv,Hui:2022vbh} admit a fully nonlinear extension at the level of the Weyl metric \eqref{eq:weyl}, which is responsible for the vanishing of the nonlinear Love numbers. We further demonstrate the relation between these nonlinear symmetries and the Geroch symmetry that is well-known to be associated with dimensional reduction to two dimensions \cite{Ehlers:1957zz,Geroch:1970nt,Breitenlohner:1986um,Maison:2000fj,Lu:2007zv,Lu:2007jc,Maison:1978es,Schwarz:1995td,Schwarz:1995af}.

The paper is organized as follows, in accordance with
four main points:
\begin{itemize}
\item An axisymmetric, static metric in general relativity takes the
  Weyl form as in \cref{eq:weyl}. The corresponding fully
  nonlinear Einstein equations imply a linear equation
  \eqref{eq:psi-eom} for $\psi$. How these come about is best
  explained by the logic of dimensional reduction. This is discussed
  in \cref{sec:dimred}.
\item Breaking $\psi$ into the Schwarzschild background $\psi_\sch$
  and perturbation $\hat \psi$ (without assuming $\hat \psi$ is
  small), the equation of motion for $\hat\psi$ can be solved to show
  that it does not have a tidal tail, i.e.,~tail being one that decays as some power of
  radius at large distances. This is discussed in
  \cref{sec:distorted}.
\item The absence of a tidal tail for $\hat\psi$ is suggestive of
  vanishing Love numbers. To confirm this is the case, we make use
  of the EFT of the black hole as a
  point object, in which Love numbers, linear as well as nonlinear,
  are clearly defined. We demonstrate the vanishing of the Love number couplings
  associated with {an infinite subset of operators} involving only even
  perturbations (but to all orders in the number of spatial derivatives and in the number of fields). This is done by matching with the Weyl solution.
  Even though the Weyl solution is axisymmetric, the vanishing of all
  the even Love number couplings goes beyond axiysmmetry.
  This is discussed in \cref{sec:EFT}.
\item The dynamics governing $\hat\psi$ has symmetries which are
  ultimately responsible for the phenomenon of having no tidal tail.
  These are explained in \cref{sec:sym}. One can think of them as
  the nonlinear generalization of the ladder symmetries discussed
  in \cite{Hui:2021vcv,Berens:2022ebl}. We connect them with
  the well-known Geroch group for dimensionally reduced spacetimes.
\end{itemize}
Points 1 and 2 are essentially known results, albeit repackaged.
Points 3 and 4 are the main new results. We conclude
in \cref{sec:discuss} with a few thoughts on interesting issues to be
explored in the future. Several appendices follow up with technical
details: \cref{app:dimred} goes over dimensional reduction,
\cref{app:reg} gives a brief overview of Geroch and Hartle's proof of
the regularity of the distorted potential, \cref{app:BHPT} connects
the Weyl gauge and the Regge--Wheeler gauge in
first order in perturbation theory,
\cref{sec:axiop} contains a discussion on a convenient way of arranging operators in the point-particle EFT, 
\cref{sec:ladders} summarizes how the
horizontal ladder symmetries and the Wronskian are related,
and \cref{app:geroch} is a primer on Geroch symmetry, focusing on
aspects relevant to our problem of interest.

\paragraph{Notation and conventions.} We adopt units such that
$c=\hbar=1$ and use the mostly plus signature for the metric,
$(-,+,+,+)$. We use both the Newtonian gravitational constant $G$ and
the reduced Planck mass $\Mp^{-2}=8\pi G$ to describe the coupling
strength of gravity.

We work at various points in $D=2,3,4$-dimensional spacetimes with metrics $g_{2,ab}$, $g_{3,\Ij}$, and $g_{4,\mn}$, respectively. To avoid confusion we reserve Greek letters $\mu,\nu,\cdots=0,1,2,3$ for 4D spacetime indices, mid-alphabet Latin letters $i,j,k,...=1,2,3$ for 3D spatial indices, and Latin letters $a,b,...=1,2$ for indices on the 2D spatial manifold obtained from reduction along the azimuthal isometry direction, e.g., $x^a=(r,\theta)$.

Objects associated to these metrics are distinguished either explicitly with subscripts or implicitly by their indices, as needed; for instance, $\nabla_i$ is the covariant derivative with respect to $g_3$, rather than the spatial component of the 4D covariant derivative $\nabla_\mu$.
We use $\epsilon_{a_1\cdots a_D}$ to refer to the totally antisymmetric Levi-Civita tensor in $D$ dimensions, in particular $\epsilon_{0123} = \sqrt{-g_4}$, $\epsilon_{123}=\sqrt{g_3}$, and $\epsilon_{12}=\sqrt{g_2}$.
In forms notation, where there are no indices, we will write the Hodge
star in $D$ dimensions as $\star_D$ and reserve the star operator for
2D, $\star\equiv\star_2$.
The symbol $\langle \, \cdots \rangle$ denotes symmetrization over the enclosed indices with subtraction of traces, e.g., $A_{\langle \mu}B_{\nu\rangle}=\frac{1}{2}(A_{\mu}B_{\nu}+ A_{\nu}B_{\mu})-\frac{1}{4}A^{\alpha}B_{\alpha}g_{\mu\nu}$.

\paragraph{Note added.} During the completion of this work, we have
become aware of a similar work by Antonio Riotto and Alex Kehagias \cite{Kehagias:2024rtz},
and have coordinated the arXiv submission. Our major conclusions,
where they overlap, agree.

\section{Dimensional reduction and the Weyl ansatz}
\label{sec:dimred}

In this paper we are principally interested in four-dimensional vacuum
spacetimes that are static and axisymmetric. This means that there is
a coordinate system $x^\mu = (t,x^a,\phi)$, with $a=1,2$, such that
$\xi=\partial_t$ and $\eta=\partial_\phi$ are Killing vectors and the
line element is invariant under $t\to-t$.\footnote{We assume $\xi$ is
  timelike and $\eta$ is spacelike, and that $\phi$ is periodic with
  period $2\pi$. Crucially we also assume $\xi$ and $\eta$ commute.}
As is well-known dating back to the work of Kaluza and Klein (KK),  if
one restricts oneself to the solution space of metrics invariant under
a given Killing vector, general relativity can be \emph{dimensionally
  reduced} to an Einstein--Maxwell--dilaton theory in one fewer
dimension. Some remarkable simplifications happen when one reduces to
three and two dimensions, resulting in the Weyl metric \eqref{eq:weyl}
and Laplace equation \eqref{eq:psi-eom}.

In this section we will
provide a brief overview of dimensional reduction with the goal of
quickly arriving at the Weyl metric and its Einstein equations. A more
detailed exposition is presented in \cref{app:dimred}.
While the discussion in this section is useful for a deeper
understanding of the Weyl construction,
for readers interested in getting to the punchline of
vanishing nonlinear Love numbers quickly,
much of it can be skipped.
The key results are:  the Weyl form of the metric, which follows from \cref{KKred}; the Laplace equation \eqref{psieom}, which follows from the Einstein
equations; and the mapping of the Schwarzschild solution between
Schwarzschild $(r,\theta)$ and Weyl
$(\rho, z)$ coordinates, in particular \cref{eq:rhortheta,eq:zrtheta,psiSch,gammaSch}.

\subsection{Reduction to Weyl and Laplace}

One can perform the KK reductions in either order, and obtain different (though of course dual) descriptions \cite{Ehlers:1957zz,Geroch:1970nt,Maison:1978es,Breitenlohner:1986um,Schwarz:1995td,Schwarz:1995af,Maison:2000fj,Lu:2007zv,Lu:2007jc}. To study Love numbers it is convenient to reduce first along $t$ and then $\phi$, parametrizing our metric as
\begin{subequations}
 \label{KKred}
\begin{align}
\dd s_4^2 &= -\mathrm{e}^{-\psi}\dd t^2+\mathrm{e}^\psi\dd s_3^2, \\
\dd s^2_3 &= \rho^2\dd \phi^2 + \mathrm{e}^{2\gamma}\dd s_2^2.
\end{align}
\end{subequations}
By construction, we exclude cross-terms involving the isometry directions,
$g_{ti}=g_{\phi a}=0$.\footnote{We thus limit ourselves to the subset
  of perturbations invariant under time reversal and $\phi$ reflection.
  Appendix \ref{app:dimred} provides more general expressions
  including these terms. It can be seen setting them to zero is a
  consistent truncation.}
The metric components are encoded in the 2D fields $\psi(x^a)$,
$\gamma(x^a)$, $\rho(x^a)$, and $g_{2,ab}(x^a)$. The four-dimensional
Einstein--Hilbert Lagrangian reduces to, up to total derivatives:
\begin{equation}
\sqrt{-g_4}  R_4 = \sqrt{g_2} \, \rho\left(R_2 +\frac{2}{\rho}\partial\rho\cdot\partial\gamma -\frac12(\partial\psi)^2 \right).\label{eq:L-2D-full-main}
\end{equation}
Indices are raised and lowered with $g_2$.
In the absence of sources we can use the equations of motion to simplify the dynamics considerably. Varying with respect to $\gamma$ we obtain an equation of motion for $\rho$,
\begin{equation}\label{eq:rho-eom}
\Box_2\rho \equiv \nabla^a\nabla_a\rho = 0.
\end{equation}
The variation with respect to $\psi$ yields its own equation of motion,
\begin{align}\label{eq:psi}
\nabla^a(\rho\nabla_a\psi)=0.
\end{align}
To compute the conformal factor $\gamma$ we project the 2D Einstein equation along $\partial_a\rho$,
\begin{equation}
\label{eq:lambda-main}
\partial_a\gamma = \frac12\rho\partial_{\langle a}\psi\partial_{b\rangle}\psi\partial^b\rho.
\end{equation}
This is a constraint equation; once we have a solution for $\psi$ we may integrate it to find $\gamma$.

With the equations of motion \eqref{eq:rho-eom}, \eqref{eq:psi}, and
\eqref{eq:lambda-main} under our belt, we now take advantage of the
fact the two-dimensional metric is conformally flat, i.e., we can find
coordinates $x^a$ for which $g_{2,ab}=f(x^a)\delta_{ab}$. In fact, we
can absorb this conformal factor $f$ into the definition of
$\mathrm{e}^{2\gamma}$ to set $g_2=\delta$.
Once this is done, the $\rho$ equation of motion \eqref{eq:rho-eom} tells us $\rho(x^a)$ is a harmonic function on $\mathbb{R}^2$. This allows us to treat $\rho$ as a \emph{coordinate} rather than a field (as long as it is not constant). This harmonic coordinate system is known as \emph{Weyl canonical coordinates},
\begin{equation}
x^a = (\rho,z),
\end{equation}
where $z$ is defined as the dual to $\rho$,\footnote{This equation is justified most easily in forms notation: \cref{eq:rho-eom} may be written $\dd\star\dd\rho=0$, from which it follows that there exists a $z$ satisfying $\dd z=-\star\dd\rho$.}
\begin{equation}\label{eq:z}
\partial_a z = \epsilon_a{}^b\partial_b\rho.
\end{equation}
Note that the 2-metric in these coordinates is then $\dd\rho^2+\dd
z^2$.\footnote{We emphasize that this construction only works in the absence of sources; if $\Box_2\rho\neq0$, then $z$ as defined in \cref{eq:z} does not exist, and we must again treat $\rho$ as a field, not a coordinate. This will be relevant in \cref{sec:EFT} when we couple gravity to a delta-function source.}

In Weyl canonical coordinates $(\rho,z)$ the $\psi$ equation of motion \eqref{eq:psi} takes the form \eqref{eq:psi-eom},\footnote{The components of \cref{eq:lambda-main} are
\begin{equation}
\partial_\rho\gamma = \frac14\rho\left((\partial_\rho\psi)^2-(\partial_z\psi)^2\right),\quad\partial_z\gamma=\frac12\rho\partial_\rho\psi\partial_z\psi.
\end{equation}}
\begin{align}
\label{psieom}
 \left(\partial_\rho^2+\frac1\rho\partial_\rho+\partial_z^2\right)\psi = 0.
\end{align}
This equation has a simple interpretation.
Treating $\rho$ and $z$ as
the radial and height coordinates, respectively, of a cylindrical
coordinate system $(\rho,\phi,z)$ in a fictitious flat 3-space,\footnote{We call it fictitious because it is \emph{not} a $t=\mathrm{const.}$ slice of the full spacetime, which in general is curved. Nevertheless it is a useful organizing principle for axisymmetric configurations, since we do not care about the $\phi$--$\phi$ metric component of the 3-space.}
this is simply the Laplace equation for
$\psi=\psi(\rho,z)$. Remarkably this equation is linear, even though
we are still working with fully nonlinear general relativity. The
linearity of the $\psi$ dynamics can be traced all the way to the beginning:
dimensional reduction yields a
quadratic action for $\psi$, cf.~\cref{eq:L-2D-full-main}.

Note that while we performed the dimensional reduction in a particular coordinate system, the fields have coordinate-invariant definitions in terms of the Killing vectors,
\begin{equation}
\xi^2=-\mathrm{e}^{-\psi},\quad \xi^2\eta^2=-\rho^2.
\end{equation}
Of course this is just because we chose the coordinates $(t,\phi)$ to be aligned with the isometry directions.

\subsection{Schwarzschild as a Weyl solution}
\label{sec:schwWeyl}

The Schwarzschild metric describing a non-rotating black hole
of mass $M$, in Schwarzschild
coordinates, is
\begin{equation}
\label{eq:sch}
\dd s^2_\sch = -f(r)\dd t^2 + \frac1{f(r)}\dd r^2+r^2(\dd\theta^2+\sin^2\theta  \dd\phi^2),
\end{equation}
where
\begin{equation}
f(r) = 1-\frac\rs r,\quad \rs = 2GM.
\end{equation}
This metric is static and axisymmetric with coordinates adapted to its temporal and azimuthal Killing vectors, so it is of the Weyl form.
Comparing to the Weyl metric we see that, reducing along $t$,\footnote{There is a dual description in which we reduce along $\phi$ before $t$. Then ${\rm e}^{-\psi} = r^2\sin^2\theta$, while $\rho$ is unchanged, as it is the determinant of the components of the vielbein in the $(t,\phi)$ directions. The Lagrangian is the same for both reductions, as is the combination $\lambda^2{\rm e}^\psi$ which appears in the 4D metric determinant $\sqrt{-g_4}$.} $\psi$ is given by
\begin{equation}
\mathrm{e}^{-\psi_\sch}=f(r),
\end{equation}
while the 3-metric 
\begin{equation}
\dd s^2_3 = \dd r^2 + \Delta(r) (\dd\theta^2+\sin^2\theta  \dd\phi^2)
\end{equation}
is the effective metric seen by a static scalar on Schwarzschild
\cite{Hui:2021vcv}.
Here we have defined
\begin{equation}
\label{defDelta}
  \Delta(r)\equiv r^2f(r)=r(r-\rs) \, .
\end{equation}
Further reducing along $\phi$ we see
\begin{equation}
\rho = \sqrt\Delta\sin\theta,\qquad \mathrm{e}^{2\gamma}\dd s^2_2 = \dd r^2 + \Delta \dd\theta^2.
\label{eq:rhortheta}
\end{equation}
To finish the dictionary between Schwarzschild coordinates and Weyl
canonical coordinates we integrate \cref{eq:z} to find
$z(r,\theta)$,\footnote{By this we mean solving the system of first-order partial differential equations $(\partial_rz,\partial_\theta z)=(\Delta^{-1/2}\partial_\theta\rho,-\Delta^{1/2}\partial_r\rho)$ obtained by computing the components of \cref{eq:z}.}
\begin{equation}
z = \frac12\Delta'\cos\theta = \left(r-\frac\rs2\right)\cos\theta.
\label{eq:zrtheta}
\end{equation}
The inverse coordinate transformation is
\begin{subequations}
\begin{align}
r&= \frac{l_+ + l_-+\rs}2,\\
\cos\theta &= \frac{l_+ - l_-}\rs,
\end{align}
\end{subequations}
where we have defined
\begin{equation}
l_\pm \equiv \sqrt{\rho^2+\left(z\pm \frac\rs2\right)^2}.
\end{equation}
If we view $(\rho,z)$ as Cartesian coordinates for the 2D plane, then the Schwarzschild coordinates are elliptic coordinates, since curves of constant $r$ are ellipses; in the 3D picture, where the Weyl coordinates are part of a cylindrical coordinate system, $(r,\theta)$ correspond to two of the prolate spheroidal coordinates.

In Weyl coordinates the horizon $r=\rs$ is a line segment of length $\rs$ on the $z$-axis, i.e., $\rho=0$ and $|z|\leq \frac\rs2$. Indeed we may interpret $\psi_\sch$ as (up to a factor of $-2$) the potential for a constant-density line mass,
\begin{equation}
 \label{psiSch}
\psi_\sch = -\ln\left(\frac{\sqrt{\rho^2+\left(z+\frac\rs2\right)^2}+\sqrt{\rho^2+\left(z-\frac\rs2\right)^2}-\rs}{\sqrt{\rho^2+\left(z+\frac\rs2\right)^2}+\sqrt{\rho^2+\left(z-\frac\rs2\right)^2}+\rs}\right).
\end{equation}
There is a multiplicative ambiguity between $g_{2,ab}$ and $\gamma_\sch$; making the conventional choice $\dd s_2^2 = \dd\rho^2+\dd z^2$, the conformal factor $\gamma$ is\footnote{We note that $\gamma$ may be computed without explicitly solving for $z(r,\theta)$. In particular, defining $\mathrm{e}^{2\gamma}(\dd\rho^2+\dd z^2) = \mathcal G_{ab} \dd x^a\dd x^b$ and noting that $\dd \rho = \partial_a\rho\,\dd x^a$ and $\dd z= -\star\dd\rho= -\epsilon^a{}_b\,\partial_a\rho\,\dd x^b$, it is straightforward to show that
\begin{equation}
\mathrm{e}^{-2\gamma} = \mathcal G^{ab} \partial_a\rho\partial_b\rho.
\end{equation}
This may be useful in cases where integrating \cref{eq:z} is less trivial than it is here.}
\begin{equation}
\label{gammaSch}
\mathrm{e}^{-2\gamma_\sch} = 1+\frac{\rs^2\sin^2\theta}{4\Delta}.
\end{equation}

\section{Distorted black holes}
\label{sec:distorted}

We are interested in a \emph{distorted} black hole \cite{Geroch:1982bv}, by which we mean one placed in a static,
external tidal environment. Focusing on $\psi$, let us split
\begin{equation}
\psi = \psi_\sch + \hat\psi \, ,
\end{equation}
where the distorted potential $\hat\psi$ is a (not necessarily small) perturbation away from
the Schwarzschild background $\psi_\sch$.\footnote{In fact $\gamma$ also receives a linear distortion, despite obeying a nonlinear equation of motion \cite{Geroch:1982bv}. We will not be concerned in this work with explicitly computing $\gamma$.} The linearity of the $\psi$
equation of motion is such that $\hat\psi$ itself obeys the same
Laplace equation \eqref{psieom}.

Using the same $(r,\theta)$ coordinates as defined by
\cref{eq:rhortheta,eq:zrtheta}, the Laplace equation
for $\hat\psi$ can be rewritten as
\begin{equation}
  \label{psieomSch}
  \partial_r (\Delta \partial_r \hat\psi ) +
  \frac1{\sin\theta}\partial_\theta\left(\sin\theta\partial_\theta
  \hat \psi \right) = 0 \, .
\end{equation}
This is exactly the same equation as that for a massless, static
scalar on a Schwarzschild background (in standard Schwarzschild coordinates), with
$\partial_\phi \hat \psi = 0$ for an axisymmetric field configuration.\footnote{The fact that this happens relies crucially on the axisymmetry
  of $\hat \psi$ (or $\psi$).
In Weyl coordinates, the Laplace equation \eqref{psieom} means $\hat\psi$
lives effectively in 3D flat space with the metric $\dd s^2 = \dd\rho^2 +
\dd z^2 + \rho^2 \dd\phi^2$. Turning to $(r,\theta,\phi)$ coordinates
as defined by eqs. (\ref{eq:rhortheta}) and (\ref{eq:zrtheta}), the
3D flat space metric takes the form
$\dd s^2 = \mathrm{e}^{-2\gamma_\sch} (\dd r^2 + \Delta \dd \theta^2) + \Delta \sin^2\theta \dd \phi^2$.
This is decidedly not
the effective 3D metric
$\dd s^2 = \dd r^2 + \Delta \dd \theta^2 + \Delta \sin^2\theta \dd \phi^2$
seen by a static scalar on a Schwarzschild
background \cite{Hui:2021vcv}, which is in fact the object we called $g_{3,\Ij}$ in the previous section.
Nonetheless, it can be checked that if $\partial_\phi \psi = 0$, the
resulting scalar equation of motion on either 3D background takes the
exact same form.
}
Thus from existing results (e.g.,~\cite{Hui:2021vcv}), we already know $\hat\psi$ cannot develop a tidal tail.
Let us review that argument.

We solve for $\hat\psi$ using separation of variables,\footnote{That a separable solution in these coordinates exists is due to standard arguments in the theory of differential equations, namely that the Laplace equation in flat 3D space is separable in prolate spheroidal coordinates \cite{Miller_1984}.}
\begin{equation}
\hat\psi(r, \theta) = \sum_{\ell} \hat\psi_\ell (r) P_\ell (\cos\theta) \, ,
\end{equation}
where $\ell$ is the angular momentum quantum number, and
$P_\ell$ is the Legendre polynomial (also known as Legendre function of the first
kind), satisfying
\begin{equation}
\frac1{\sin\theta}\partial_\theta\left(\sin\theta\partial_\theta
  P_\ell  
  \right) = -\ell(\ell+1)P_\ell  \, .
\end{equation}
Interestingly, writing the Laplace operator in $(r,\theta)$ coordinates we find that $\hat \psi_\ell$ also obeys a Legendre equation, with the role of $\cos\theta$ replaced by
$\Delta'/\rs = (2r/\rs) - 1$. The most general solution is thus
\begin{equation}
\hat\psi (r,\theta) = \sum_{\ell} \left[ a_\ell P_\ell\left( {\frac{2r}\rs} -
  1\right) + b_\ell Q_\ell \left({\frac{2r}\rs} -
  1\right) \right] P_\ell (\cos\theta) ,
\end{equation}
where $a_\ell , b_\ell$ are constant coefficients, and $P_\ell$ and
$Q_\ell$ are Legendre functions of the first and second kind,
respectively.
The asymptotics of $P_\ell$ and $Q_\ell$ are as follows:
\begin{subequations}
\begin{align}
P_\ell \left( {\frac{2r}\rs} -
  1\right)
  \overset{r\to\infty}{\longrightarrow}\frac{(2\ell-1)!!}{\ell!} \left( {\frac{2r}\rs} -
    1\right)^\ell  &, \quad
P_\ell \left( {\frac{2r}\rs} -
  1\right) \overset{r\to\rs^+}{\longrightarrow}1 \, ,
  \\
Q_\ell \left( {\frac{2r}\rs} -
  1\right) \overset{r\to\infty}{\longrightarrow}\frac{\ell!}{(2\ell+1)!!}
           \left( {\frac{2r}\rs} -
           1\right)^{-\ell-1} &, \quad
Q_\ell \left( {\frac{2r}\rs} -
  1\right) \overset{r\to\rs^+}{\longrightarrow}-\frac12\ln\frac{2(r-\rs)}{\rs}.
\end{align}
\end{subequations}
It is worth emphasizing that $P_\ell \left( {\frac{2r}\rs} -
  1\right)$ is a polynomial with only non-negative powers of $r$
(and dominated by $r^\ell$ at large $r$), 
while $Q_\ell \left( {\frac{2r}\rs} -
  1\right)$ does have negative powers of $r$ at large $r$.
In other words, it is only $Q_\ell$ that contains the tidal tail,
going as $1/r^{\ell+1}$
at large $r$.
Intuitively one can think of $P_\ell$ as the external tidal field,
and $Q_\ell$ as the tidal response of the object in question, in this
case the black hole.\footnote{It is worth noting that the Schwarzschild background corresponds to the monopolar decaying solution with $b_0=2$,
    \begin{equation}
  Q_0\left(\frac{2r}\rs-1\right) = \frac12\ln\frac r{r-\rs} = \frac12\psi_\sch.
  \end{equation}
By construction this term is excluded from $\hat\psi$.
}
  
At this point, we can invoke a result due to Geroch and Hartle, namely that $\hat \psi$ must be smooth on the
horizon \cite{Geroch:1982bv}. Put in another way, what they showed is that as one
approaches the horizon, the
full $\psi = \psi_\sch + \hat \psi$ for a distorted black hole
must give rise to exactly the same distributional singularity as that for a Schwarzschild
black hole. Thus, the (not necessarily small) perturbation $\hat\psi$
must be regular on the horizon. A brief summary of their derivation is given
in appendix \ref{app:reg}.

With Geroch and Hartle's result in hand, we can discard the
$Q_\ell$ solutions which diverge logarithmically at the horizon, and
thus declare\footnote{There is a  more direct  way
  of getting the condition $b_\ell = 0$ in  cases where $\hat{\psi}$
  is a small perturbation to $\psi_\sch$ (as we
  will assume  in the next section when matching to the EFT). For small deviations
  from Schwarzschild, one can expand the metric in powers of the
  distorted potential $\hat\psi$. In particular, the linearized
  $\delta g_{tt}$ component of the metric perturbation will simply be
  $\delta g_{tt}=f(r)\hat{\psi}$. This can be related to the
  linearized metric in Regge--Wheeler gauge via a suitable change of
  coordinates (see \cref{app:BHPT} for more details on the connection
  to  standard black hole perturbation theory). 
In particular, notice that, at leading order in the large-distance limit $(\rs/r\rightarrow0)$, $\delta g_{tt}$ becomes gauge invariant, i.e.~the linearized $\hat\psi$ must coincide with the Regge--Wheeler $H_0$~\cite{Regge:1957td} (see, e.g., \cref{eq:FieldsMatch}). Since the physical solution for $H_0$ does not contain any decaying falloff at large $r$ (see, e.g., \cite{Iteanu:2024dvx}), this implies  $b_\ell = 0$ for $\hat\psi$.} 
\begin{equation}
  b_\ell = 0  ,
  \label{eq:blzero} 
\end{equation}
so that $\hat \psi$ cannot have a tidal tail in $r$,\footnote{Note that we begin the sum at $\ell=2$. The monopole term $a_0$ corresponds to an unphysical constant, which can be absorbed into a constant rescaling of the coordinates. In perturbation theory, it is easy to understand that the dipole  $a_1$ does not correspond to a physical mode by comparing $\hat{\psi}$ with the linearized $\delta g_{tt}$ metric perturbation in the standard Regge--Wheeler gauge~\cite{Regge:1957td}; see also \cref{app:BHPT}.}
\begin{equation}
\hat\psi (r,\theta) = \sum_{\ell=2}^\infty  a_\ell P_\ell\left( \frac{2r}\rs -
  1\right)  P_\ell (\cos\theta) .
\label{eq:hatpsiphysicalsolution}
\end{equation}
This is a remarkable result, given that this statement comes from
solving the $\hat\psi$ equation of motion, which originates
from the fully nonlinear Einstein equation, i.e.,~without assuming
perturbations from the Schwarzschild background are small.

The absence of a decaying term at $r=\infty$ is commonly associated with vanishing Love numbers, as indeed would be the case in Newtonian gravity, where the Love numbers were originally defined. Nevertheless it is too quick to declare the vanishing of the (linear or nonlinear) Love numbers just on the basis of $\hat\psi$ lacking a tail. This is principally for two reasons. First, the identification of Love numbers with the coefficients of tail terms in $\hat\psi$ is not coordinate invariant. Second, we would ideally like a physically well-motivated definition of Love numbers, in the sense that we want to focus on an object's intrinsic static response and exclude any other effects that may contribute to its gravitational field. These problems are ameliorated by defining the Love numbers as Wilson coefficients in the worldline effective theory. In the next section we construct the relevant EFT and perform the matching to confirm our intuition that the nonlinear Love numbers vanish.

Before we do so, let us close with the observation that
we could have derived the no-tidal-tail statement in a variety of coordinate systems.
In writing down \cref{psieomSch},
we chose to go from the $(\rho, z)$ Weyl coordinates to the
$(r,\theta)$ Schwarzschild coordinates (related by 
eqs.~\eqref{eq:rhortheta} and \eqref{eq:zrtheta}).
This was a reasonable choice, both because the background
$\psi_\sch$ is typically written in $(r,\theta)$ coordinates, and
because the resulting \cref{psieomSch} takes a familiar form,
exactly the same as that for a massless static scalar around a black hole.
But as far as deriving the statement of no-tidal-tail goes,
we could have chosen more mundane coordinates
$({\mathcal R}, \vartheta)$, where
$\rho = {\mathcal R} \sin\vartheta$ and
$z =  {\mathcal R} \sin\vartheta$. After all, as remarked
before, \cref{eq:psi-eom} tells us that
$\psi$ (or $\hat \psi$) effectively lives in 3D flat space. Switching
to ordinary spherical coordinates $({\mathcal R}, \vartheta)$
would have allowed us to conclude $\hat\psi$ has no tidal tail as
well.\footnote{In fact, the argument is simpler, because the regular and irregular radial
solutions are purely power-law: $\mathcal R^\ell$ and $1/\mathcal
R^{\ell+1}$.}
We will have more to say about this in \cref{sec:sym}.
Note however that the Schwarzschild background $\psi_\mathrm{sch}$ is
a bit more complicated in these coordinates; rather than corresponding
to a pure monopole $b_0$, in spherical coordinates it corresponds to
turning on all even multipoles,
$b_{r,\mathrm{even}}=2(GM)^{r+1}/(r+1)$. These are familiar as the
Weyl multipole moments of Schwarzschild (see e.g.~\cite{Gurlebeck:2015xpa}).

\section{Matching to the worldline effective field theory}
\label{sec:EFT}

\subsection{General considerations and nonlinear Love numbers}

A rigorous definition of Love numbers in general relativity, which allows one to address ambiguities associated to gauge freedom in the theory, and provides a systematic framework to incorporate nonlinear response, is in terms of the worldline EFT \cite{Goldberger:2004jt,Goldberger:2005cd,Goldberger:2006bd} (see, e.g., \cite{Foffa:2013qca,Rothstein:2014sra,Porto:2016pyg,Levi:2018nxp,Goldberger:2022ebt,Goldberger:2022rqf} for some reviews). The EFT formalizes the simple intuition that any object from far away looks, in first approximation, like a point particle. Finite-size effects are then captured in terms of higher-dimensional operators attached to the worldline, organized in a derivative expansion. In this language, the Love numbers correspond to particular Wilson coefficients in the EFT. Such a definition has the added bonus of equipping the Love numbers with a clear physical interpretation, as the effective theory by construction separates the object's intrinsic response from other effects.

The  point-particle action along the black hole's worldline is 
\begin{align}
S_\mathrm{p.p.} &= -M\int\dd \tau = -M\int \dd \lambda\sqrt{-g_\mn \frac{\dd x^\mu}{\dd\lambda}\frac{\dd x^\nu}{\dd\lambda}} \label{eq:PP}\\
& = \frac12\int\dd\lambda \left(e^{-1}g_\mn \frac{\dd x^\mu}{\dd\lambda}\frac{\dd x^\nu}{\dd\lambda}-M^2e\right),\label{eq:PP-ein}
\end{align}
where $\tau$ is the proper time along the worldline, which we parametrize as $x^\mu(\lambda)$ for an arbitrary affine parameter $\lambda$, $e$ is an einbein which can be integrated in or out as convenient, and $M$ is the mass. 
The goal is to study the induced response, which we extract  by probing the object with some external tidal field. The latter  solves  the vacuum Einstein equations, obtained from  the bulk Einstein--Hilbert action,
\begin{align}
S_\mathrm{EH} &= \frac{\Mp^2}2\int\dd^4x\sqrt{-g_4}R_4 \nonumber\\
&= -\frac{\Mp^2}4\int\dd t \,\dd\phi\, \dd^2x\sqrt{g_2}\rho(\partial\psi)^2,
\end{align}
with $\Mp$ the Planck mass. The tidal deformation and the multipole moments it induces %
can  be described by adding an action term $S_\mathrm{int}$, which parametrizes the interaction between the external gravitational field and the object. The final action is therefore
\begin{equation}
\label{eq:S-EFT}
S = S_\mathrm{EH} + S_\mathrm{p.p.} + S_\mathrm{int} . 
\end{equation}
At quadratic order, the interaction term $S_\mathrm{int}$ can be written as
\begin{equation}
S_\mathrm{int} = \sum_{\ell=2}^\infty  \int \dd \tau \left( 
Q_E^{\mu_L} E_{\mu_L} + Q_B^{\mu_L} B_{\mu_L} \right)+ \dots ,
\label{Sint}
\end{equation}
where ellipses stand for nonlinear couplings between $Q_{E,B}$ and powers of $E$ and $B$, and
where we introduced the multi-index notation $\mu_L\equiv \mu_1\cdots \mu_\ell$. Here we have defined 
\begin{align}
E_{\mu_1\cdots\mu_\ell}&  \equiv P_{\langle \mu_1}^{\nu_1} \cdots P_{\mu_{\ell-2} \vert }^{\nu_{\ell-2}} \nabla_{\nu_1} \cdots \nabla_{\nu_{\ell-2} } E_{\vert \mu_{\ell-1}\mu_\ell \rangle},
\\
B_{\mu_1\cdots\mu_\ell} & \equiv P_{\langle \mu_1}^{\nu_1} \cdots P_{\mu_{\ell-2} \vert }^{\nu_{\ell-2}} \nabla_{\nu_1} \cdots \nabla_{\nu_{\ell-2} } B_{\vert \mu_{\ell-1}\mu_\ell \rangle} ,
\end{align}
where {
\begin{equation}
E_{\mu\nu} \equiv - R_{\rho\langle \mu\nu\rangle \sigma}u^\rho u^\sigma,
\qquad
B_{\mu\nu} \equiv \frac{1}{2} {\epsilon_{\gamma \langle \mu}}^{\alpha\beta} R_{\nu\rangle \delta \alpha\beta} u^\delta u^\gamma 
\label{EmunuBmunu} 
\end{equation}
correspond to the  electric and magnetic parts of the four-dimensional Riemann tensor $R_{\mu\rho\nu\sigma}$, respectively,\footnote{The $E$ and $B$ fields are often defined in the literature in terms of the Weyl tensor, instead of the Riemann tensor. The two choices are completely equivalent, up to a field redefinition  in the effective theory.} with $u^\mu = \dd x^\mu /\dd \tau$  the particle's four-velocity, %
while $P^\nu_\mu$ is the projector on the plane orthogonal to $u^\mu$, i.e.,
\begin{equation}
P^\nu_\mu \equiv \delta^\nu_\mu + u^\nu u_\mu .
\end{equation}
In the following, we will work in  the rest frame of the point
particle (aligned, by construction, with the timelike Killing
direction), where  $u^\mu=(1,\vec{0})=\xi^\mu$. In this frame,  only
the spatial components of $E_{\mu\nu}$ and $B_{\mu\nu}$ are
non-vanishing.}
{Note that the notation $\langle \mu_1 \mu_2 \cdots \rangle$ denotes
symmetrization with all traces removed.}

In \cref{Sint}, $Q_E^{\mu_L}$ and $Q_B^{\mu_L}$ correspond to the  deformation of the body induced by the coupling to the external   $E_{\mu_L}$ and $B_{\mu_L}$ fields. Note that the action \eqref{Sint} can be used  to describe not only the body's conservative response, but also finite-size dissipative effects. Dissipation can be incorporated by interpreting    $Q_E^{\mu_L}$ and $Q_B^{\mu_L}$ as composite operators that depend on additional unknown gapless  degrees of freedom on the worldline, which are responsible, e.g.,~for absorption~\cite{Goldberger:2005cd,Goldberger:2020fot,Ivanov:2022hlo}.
Since we are interested in performing the matching with the nonlinear solution \eqref{eq:hatpsiphysicalsolution} for static and axisymmetric spacetimes,  we will restrict ourselves in what follows to the conservative, even-parity  sector. In other words, we will focus  on operators that contain the $E_{\mu_L}$ field only, and we will neglect dissipation, which is absent for Schwarzschild black holes in the static regime. 

The idea is to solve for $Q_E$ in response theory and plug its solution back into the action \eqref{Sint}. In general, we shall parametrize the one-point function of $Q_E$ as~\cite{Iteanu:2024dvx}
\begin{equation}
\langle Q_E^{i_L}(\tau) \rangle 
= \sum_{n=1}^\infty \int \dd \tau_1 \cdots \int \dd\tau_n \, {}^{(n)} \! {\cal R}^{i_L\vert i_{L_1} \cdots i_{L_n}}(\tau-\tau_1,\dots,\tau-\tau_n) E_{i_{L_1}}(\tau_1) \cdots E_{i_{L_n}}(\tau_n) ,
\label{QREn}
\end{equation}
where $i_L\equiv i_1 \cdots i_\ell$, and  where ${}^{(n)} \! {\cal R}$ is the $n^{\text{th}}$-order response function.\footnote{Note that the right-hand side of \cref{QREn} contains only powers of $E$ because, as alluded before, we are not considering the odd sector. In general, $\langle Q_E^{i_L}(\tau) \rangle$ can be sourced  also by $B$ fields at nonlinear order~\cite{Bern:2020uwk,Riva:2023rcm,Hadad:2024lsf,Iteanu:2024dvx}.} Note that \cref{QREn} parallels the expression of the induced electric polarization of a medium in nonlinear response theory of nonlinear optics \cite{10.1093/acprof:oso/9780198702764.001.0001}; in fact, the problem  that we are  trying to solve here is conceptually very similar to  electromagnetism, where the induced polarization  is written  as an expansion in powers of the applied optical electric field strength, parametrized in terms of a general polarization response function. By construction, the tensor ${}^{(n)} \!{\cal R}^{i_L\vert i_{L_1} \cdots i_{L_n}}$ is symmetric under the exchange of its last $n$ multi-indices $i_{L_a}$. In addition, causality implies that ${}^{(n)} \! {\cal R}$ vanishes whenever any of its arguments $\tau-\tau_j$, for $j=1,\dots,n$, is negative. In the absence of dissipation and under the assumption of  static tides, the time dependence in ${}^{(n)} \! {\cal R}$ factors out as the product of Dirac deltas, i.e.,~${}^{(n)} \! {\cal R}
\propto   \delta(\tau-\tau_1)\cdots  \delta(\tau-\tau_n)$.\footnote{This can be thought of as the leading order of an adiabatic approximation~\cite{Goldberger:2005cd,Goldberger:2020fot,Ivanov:2022hlo,Saketh:2023bul}. More generally, in the presence of finite-frequency effects, ${}^{(n)} \! {\cal R}$ will contain also time derivatives of $\delta(\tau-\tau_j)$.}
In addition, for non-rotating objects, its tensorial structure boils down to  the tensor product of Kronecker deltas. For instance, at leading order in the number of spatial derivatives, one has~\cite{Bern:2020uwk,Riva:2023rcm,Iteanu:2024dvx}
\begin{equation}
{}^{(n)}\!{\cal R}^{ij\vert i_1 j_1 \cdots i_n j_n}
=  \sum_k \lambda_k \, {\rm perm}_k \left[ \delta^{\langle i}_{\vert j_n \rangle} \delta^{j \rangle}_{\langle i_1 } \delta^{\langle i_2}_{j_1 \rangle} \delta^{ j_2\rangle}_{\langle i_3}   \cdots \delta^{j_{n-1} \rangle }_{\langle i_n\vert} \right]  \delta (\tau-\tau_1)\cdots  \delta (\tau-\tau_n) ,
\label{eq:defRnth}
\end{equation}
with some coefficients $\lambda_k$,
and similarly at higher orders. In \cref{eq:defRnth} we are summing over permutations of the enclosed  indices.
 Plugging the solution \eqref{QREn} into the action \eqref{Sint}
 including nonlinear couplings, one then finds a series of effective
 interaction terms in the form of local operators involving
 contractions of $E_{ij}$ and derivatives thereof. In symbols,{
\begin{align}
S_\mathrm{int} & = \displaystyle
\sum_{n=1}^\infty
\int \dd\tau \sum_{\mathclap{\substack{\ell, \ell_1,  \cdots , \ell_n \\ \ell=\ell_1 \otimes \cdots \otimes \ell_n }}}
  F\left( \lambda_{\ell\ell_1\cdots\ell_n}^{(n)} E_{ i_L} E_{ i_{L_1}} \cdots E_{ i_{L_n}}\right) 
\\
& = \displaystyle \int \dd\tau \Bigg[ \sum_{\mathclap{\substack{\ell,
     \ell_1\\ \ell=\ell_1}}}
     \lambda_{\ell\ell_1}^{(1)}E_{i_L} E^{i_{L_1}} +
     \sum_{\mathclap{\substack{\ell, \ell_1, \ell_2 \\ \vert \ell_2-
  \ell_1 \vert  \leq \ell \leq \ell_1+\ell_2 }}}  F \left( \lambda_{\ell\ell_1\ell_2}^{(2)}  E_{i_L} E_{i_{L_1}}E_{i_{L_2}}\right)  +\dots
 \Bigg],
\label{Sintexp}
\end{align}
where $F(\, \cdots )$ is responsible for all possible contractions
among the indices of the enclosed tensor, and
$\lambda_{\ell\ell_1\cdots\ell_n}^{(n)}$ are the (nonlinear) Love numbers
coupling encoding information about the conservative tidal
deformability of the compact object, including nonlinearities.
Note that the first set of operators involving two $E$'s are the standard ones dictating linear response. The corresponding coupling
$\lambda^{(1)}_{\ell\ell_1}$ might look a bit unfamiliar, but represents none
other than what is usually labeled as $\lambda^{(1)}_\ell$. (Note that
the sum over $\ell$ and $\ell_1$ reduces to a sum over $\ell$ only,
under the condition $\ell_1 = \ell$. We will thus
sometimes refer to this coupling as
$\lambda^{(1)}_{\ell\ell}$, or simply as $\lambda^{(1)}_\ell$.)
This is as it should be, for
$E_{i_L} E^{i_{L_1}}$ represents the contraction between
$E_{i_1 i_2 ... i_\ell}$ and $E_{j_1 j_2 ... j_{\ell_1}}$, keeping in
mind that $E$ is completely trace-free (i.e.~the contraction of indices
within a single $E$ is not allowed), and so one must have $\ell =
\ell_1$ in order to have all indices contracted properly.
As a result, for each multipole, linear response is fully captured by a single coefficient, $\lambda^{(1)}_\ell$.
Quadratic response involves 
three $E$'s, contracting indices among
$E_{i_1 i_2 ... i_\ell}$, $E_{j_1 j_2 ... j_{\ell_1}}$ and $E_{k_1 k_2
    ... k_{\ell_2}}$. The contraction of indices among the last two $E$'s
  gives rise to a number of possibilities. One could have an object
  with $\ell_1 + \ell_2$ free indices (i.e.~no contraction at all). In
  that case, a successful contraction with the first $E$ requires
  $\ell = \ell_1 + \ell_2$. The contraction of indices among the last
  two $E$'s can also result in as few as $|\ell_2 - \ell_1|$ free
  indices, in which case the first $E$ must carry
  $\ell = |\ell_2 - \ell_1|$ indices. One recognizes this is exactly
  the selection rule associated with the addition of angular momentum \cite{Riva:2023rcm,DeLuca:2023mio}.
At this perturbative order, it is easy to realize that, similarly to linear response, there can only be at most one single nontrivial independent contraction of indices, and thus one single $\lambda_{\ell\ell_1\ell_2}^{(2)}$, for given multiplet $(\ell\ell_1\ell_2)$.\footnote{\label{foot2nd}To see this, consider a generic cubic operator  $E_{i_1\cdots i_l}E_{j_1\cdots j_m}E_{k_1\cdots k_n}$, where we shall assume in full generality that $l\leq m\leq n$. Indices are contracted with some constant tensor. One shall proceed constructively and assume that for fixed $l$, $m$ and $n$ one such contraction exists and is nontrivial. Then, it is not hard to convince ourselves that it must be unique. Indeed, any other combination, which does not change  $l$, $m$ and $n$, must necessarily involve a trace over at least a couple of indices in one of the $E$'s. As an example, consider the operator $E_{i_1i_2}E_{j_1j_2}E_{k_1k_2k_3k_4}$  with $l=m=2$, $n=4$. The only nontrivial contraction is $E_{i_1i_2}E_{j_1j_2}E^{i_1i_2j_1j_2}$: one cannot contract the $E$'s differently without inevitably tracing over two indices of $E_{k_1k_2k_3k_4}$. The only other option is to change $l$, $m$ and $n$: e.g., one can have $E_{i_1i_2}{E^{i_1}}_{j_1j_2}E^{i_2j_1j_2}$ which enters at the same order in the derivative expansion but differs in how derivatives are arranged on the fields. We stress that we are not saying that at given order in the number of derivatives in the EFT there is only one independent operator. More operators can be present at fixed $l+m+n$, but these will necessarily differ by the values of  the numbers $l$, $m$ and $n$; as such, as far as the matching is concerned, the couplings of these operators can be probed independently by considering different harmonic configurations for the tidal and response fields.}
 The pattern repeats at subsequent levels. 
For instance, the cubic response operators will involve four $E$'s, involving $\ell, \ell_1, \ell_2, \ell_3$, obeying some particular constraint among them. It is worth noting that the sum of all the $\ell$'s, i.e.~$\ell + \ell_1 + \ell_2 +\cdots$, must be even.
Note that starting from $\mathcal{O}(E^4)$, for a fixed set of angular momenta $(\ell\ell_1\cdots
\ell_n)$, multiple $\lambda$'s can appear.
The exact counting of inequivalent contractions and independent Wilson couplings  can be obtained from standard group theory arguments~\cite{Bern:2020uwk,Haddad:2020que,Ruhdorfer:2019qmk}. In the following we will not attempt to provide a classification, which we leave for future work. Instead, we will show that, after a suitable basis choice  in the EFT,  matching to explicit  solutions of static and axisymmetric black hole perturbations in general relativity sets to zero one coupling for each distinct multiplet. At those orders where one single independent coupling is present, the matching is exhaustive and fixes all the freedom. When instead more inequivalent contractions of operators are present, some Love number couplings are left unconstrained (fixing them requires going beyond axisymmetry).
}

{Anticipating that we will perform the matching for axisymmetric tidal fields, let us reorganize the effective expansion \eqref{Sintexp} so that, at each perturbative order and for a fixed multiplet, all but one operator yield a vanishing contribution when computed in axisymmetric configurations. In practice, we shall  rewrite \eqref{Sintexp} in full generality as
  \begin{equation}
  S_\mathrm{int} = \displaystyle \sum_{n=1}^\infty
\int \dd\tau \sum_{\mathclap{\substack{\ell ,\ell_1 , \cdots , \ell_n \\ \ell=\ell_1 \otimes \cdots \otimes \ell_n }}}
  \lambda_{\ell\ell_1\cdots\ell_n}^{(n)} E_{ i_L} E_{ i_{L_1}} \cdots
  E_{ i_{L_n}} + \tilde S_\mathrm{int}\, ,
\label{Sintalt}
\end{equation}
where $E_{ i_L} E_{ i_{L_1}} \cdots
  E_{ i_{L_n}}$ represents a generic (nontrivial) contraction, while $\tilde S_\mathrm{int}$ consists of all the Love number operators that vanish in the Einstein equations for
axisymmetric configurations, whose couplings we will uniformly refer
to as $\tilde \lambda$'s.
As we explain in appendix~\ref{sec:axiop}, one can always choose an operators basis in the EFT in such a way to extract $\tilde S_\mathrm{int}$.
In the matching below with the Weyl solution, we will not be able to constrain the $\tilde\lambda$'s. We will however be able to constrain
the $\lambda$'s, because they multiply Love number operators that
do not vanish under axisymmetry. In \eqref{Sintalt}, 
we keep the contraction among the
indices implicit: the idea is that we can choose any one
of the possible contractions, {\it but only one}. All the other
contractions can be combined in such a way that they fall in the class
represented by $\tilde S_{\rm int}$.  }

\subsection{Matching to Weyl solution}
\label{sec:matchweyl}

{Let us now concretely see the previous logic in action.
The goal is to extract the explicit values of the effective Love number coefficients $\lambda_{\ell\ell_1\cdots\ell_n}^{(n)}$ in \eqref{Sintalt},  in the case of Schwarzschild black holes.
To this end, we will compute the one-point function of the response field in the EFT \eqref{eq:S-EFT} with the interaction terms \eqref{Sintalt}, and  compare it with  the full solution \eqref{eq:hatpsiphysicalsolution} in general relativity. To perform this matching, we find it convenient to use in the EFT calculation the same Weyl coordinates that we used above; this will save us from performing extra gauge transformations, or constructing nonlinear gauge-invariant quantities \cite{Hui:2020xxx,Riva:2023rcm,Iteanu:2024dvx}.
We will proceed iteratively and show that, order by order in perturbation theory, the inhomogeneous solution for the response field in Minkowski space has purely decaying falloff, which is absent in the large-distance expansion of \cref{eq:hatpsiphysicalsolution}. Hence, 
the coefficients $\lambda_{\ell\ell_1\cdots\ell_n}^{(n)}$ in \eqref{Sintalt} vanish. }

Let us start by considering a static, axisymmetric distortion of  flat spacetime in spherical coordinates, $(t,r,\theta,\phi)$.
Recall that a crucial ingredient in the reduction of the Weyl metric to the form \eqref{eq:weyl} was the harmonic condition $\Box_2\rho=0$, which allowed us to choose $\rho$ as a coordinate in the metric. 
In the presence of the source $S_\mathrm{p.p.}$ and the interactions $S_\mathrm{int}$, this is no longer necessarily true.  We should therefore treat the  $g_{\phi\phi}$ component of the metric as an independent field. Concretely, we will use here the following ansatz:
\begin{equation}\label{eq:weylflat}
\dd s^2_4 = -\mathrm{e}^{- \psi}\dd t^2 + \mathrm{e}^\psi\left[\mathrm{e}^{2 \gamma}(\dd r^2+r^2\dd \theta^2)+ \rho^2\dd\phi^2\right].
\end{equation}
We stress that here $\rho$ is not a coordinate; we will interpret it as a field, depending on $r$ and $\theta$, like  $\psi$ and $\gamma$. In the limit of vanishing source and interactions, one can use the Einstein equations to set $\rho = r\sin\theta$, which, together with $z = r\cos\theta$, would recover \cref{eq:weyl}.\footnote{Although we use the same symbols, to be precise the coordinates $r$ and $\theta$ here are different from the ones used in \cref{sec:schwWeyl}. They coincide only for $M=0$. Since we will be working here at leading order in the flat-space limit, this difference is immaterial. It would be  important instead if one tried to perform the matching at subleading order in $M$, or reconstruct the nonlinear tidal field solution from the EFT \cite{Riva:2023rcm,Iteanu:2024dvx}.} 
Later on we will argue that $\rho = r\sin\theta$ actually remains a legitimate choice in the EFT \eqref{eq:S-EFT}, even in the presence of $S_\mathrm{int}$.

The EFT is defined in the infrared, where spacetime is close to Minkowski ($\psi=0=\gamma$), so we can identify $\psi=\hat\psi$  and $\gamma=\hat{\gamma}$, and use them as expansion parameters. Herein we will use $\psi$ and $\hat\psi$, and $\gamma$ and $\hat\gamma$, interchangeably when doing perturbation theory around flat space.
Each field can be separated into tidal and response contributions, e.g.,
\begin{equation}
\psi = \psi_{\text{tidal}} + \psi_{\text{resp}}, 
\label{tidalresponsesplit}
\end{equation}
and analogously for $\gamma$ and $\rho^2$. 
Note that the split \eqref{tidalresponsesplit} is non-ambiguous. In fact, from the Einstein equations obtained from the action $S_\mathrm{EH}  + S_\mathrm{int}$,\footnote{In order to study the induced response and perform the matching at leading order in the $M/r\rightarrow0$ limit, we can safely neglect the point-particle contribution $S_\mathrm{p.p.}$ in the EFT \eqref{eq:S-EFT}. $S_\mathrm{p.p.}$ would be relevant, for instance, to study $\mathcal{O}(\rs)$ corrections to the tidal solution, or to reconstruct the Schwarzschild background metric~\cite{Goldberger:2004jt}.} which are schematically
\begin{equation}
G_{\mu\nu}
= - \frac{2}{\Mp^2}\frac{\delta}{\delta g_{\mu\nu}} S_\mathrm{int} ,
\label{EinEqsint}
\end{equation}
where $G_{\mu\nu}$ is the Einstein tensor $G_{\mu\nu}\equiv R_{\mu\nu}-\frac{1}{2}R g_{\mu\nu} $, $\psi_{\text{tidal}}$ is defined as the solution to the homogeneous equation in the bulk, $G_{\mu\nu}=0$, i.e.,~in the absence of $S_\mathrm{int}$. In particular, following the considerations above, from the combination
\begin{equation}
\rho \, \mathrm{e}^{2\gamma}( \mathrm{e}^{2\psi} R_{tt}-\rho^{-2}R_{\phi\phi} )= \left(\partial_r^2 +\frac{1}{r}\partial_r + \frac{1}{r^2} \partial_\theta^2 \right)\rho=0 ,
\label{RttRpp}
\end{equation}
which is valid in the bulk,
one can fix in full generality   $\rho= r \sin\theta$, and find
\begin{equation}
\psi_{\text{tidal}} =\sum_\ell c_\ell r^\ell P_\ell(\cos\theta) .
\label{psitidalflat}
\end{equation}

Note that, in the perturbative regime $\psi_{\text{tidal}}\ll 1$, one recovers at linear order the usual gravitational tidal field that grows as $r^\ell$ at large $r$ and that is regular at $r=0$. Given the tidal profile \eqref{psitidalflat}, we shall then add the response piece $\psi_{\text{resp}}$ which by definition solves the inhomogeneous equations of motion, in the presence of the interaction term $S_\mathrm{int}$. Note that, even though we introduced the response field at the level of the nonlinear field $\psi$ in \cref{tidalresponsesplit}, we are going to expand in  $\psi_{\text{resp}}\ll 1$ and perform the matching perturbatively order by order in the number of fields. This will ensure that the right-hand side of \cref{EinEqsint} contains the tidal field only, while $\psi_{\text{resp}}$ will appear only in $G_{\mu\nu}$, as we will see explicitly.

\paragraph{Linear response.} As a warm-up, let us start by performing the matching at linear order. We will follow \cite{Hui:2020xxx,Rai:2024lho}. As a consistency check, we will show that the matching recovers the standard result of vanishing linear Love numbers \cite{Rothstein:2014sra}. We will then show how to systematically extend the result to higher orders in perturbation theory.

{At second order in the worldline EFT, a minimal set of independent quadratic operators is given by $E_{i_L} E^{i_L}$ in \cref{Sintexp}. 
The Einstein equations \eqref{EinEqsint}, with linearized $S_\mathrm{int}$, read}
\begin{equation}
G_{\mu\nu}^{(1)}
= - \frac{1}{\Mp^2} \sum_{\ell=2}^\infty (-1)^\ell   \lambda_{\ell}^{(1)}\partial^{i_1} \cdots \partial^{i_\ell}
\left( \delta^{(3)}(\vec{x}) \,  \partial_{i_1} \cdots \partial_{i_\ell} \psi_{\text{tidal}} \right) \delta_\mu^0 \delta_\nu^0.
\label{GmunuLinear}
\end{equation}
{Here we have expanded the electric part of the curvature tensor to linear
order, 
\begin{equation}
E_\Ij = -\frac12\partial_{\langle i}\partial_{j \rangle}\psi +\mathcal{O}(\psi^2, \psi\gamma),
\label{Eijexp}
\end{equation}
replaced $\psi$ with $\psi_{\text{tidal}}$ on the right-hand side, and placed the point particle at the origin of the coordinate system.}
Although some of the manipulations below become more straightforward in  Weyl coordinates by working with the full Einstein tensor  $G_{\mu\nu}$, we  put a superscript in $G_{\mu\nu}^{(1)}$ to recall  that \cref{GmunuLinear} is valid only at linear order. 
 Taking a linear combination with the trace, it is convenient to rewrite \cref{GmunuLinear} as
\begin{equation}
R_{\mu\nu}^{(1)}
= - \frac{1}{\Mp^2} \sum_{\ell=2}^\infty (-1)^\ell   \lambda_{\ell}^{(1)}\partial^{i_1} \cdots \partial^{i_\ell}
\left( \delta^{(3)}(\vec{x}) \,  \partial_{i_1} \cdots \partial_{i_\ell} \psi_{\text{tidal}} \right) \left( \delta_\mu^0 \delta_\nu^0  +\frac{1}{2}\eta_{\mu\nu}  \right) \equiv J_{\mu\nu}^{(1)} ,
\label{GmunuLinear2}
\end{equation}
Note that, since $J^{(1)}_{tt}-J^{(1)}_{\phi\phi}/(r\sin\theta)^2=0$, from \cref{RttRpp} it follows that we can still  choose at linear order $\rho=r\sin\theta$, despite the presence of the interaction term to the worldline.\footnote{Recall that we are working on flat spacetime at linear order in perturbation theory. This means for instance that we can, at this order, approximate to unity the exponential factors appearing in the combination \eqref{RttRpp}, which give contributions at higher orders.}  Corrections, if present, will start from second order in the fields $\psi$ and $\gamma$. Hence, fixing $\rho=r\sin\theta+ \mathcal{O}(\psi^2,\psi\gamma)$ and focusing on the $t$--$t$ component of \cref{GmunuLinear2}, one finds the equation
\begin{equation}
\vec{\nabla}^2\psi_{\text{resp}} = \frac{1}{\Mp^2} \sum_{\ell=2}^\infty (-1)^\ell   \lambda_{\ell}^{(1)}\partial^{i_1} \cdots \partial^{i_\ell}
\left( \delta^{(3)}(\vec{x}) \,  \partial_{i_1} \cdots \partial_{i_\ell} \psi_{\text{tidal}} \right)  ,
\label{eqpsilin1}
\end{equation}
where $\vec{\nabla}^2$ is the Laplace operator in flat space, and where   we have kept only terms linear in the fields. At this order, the equation does not contain $\gamma$, and we can readily solve  for $\psi_{\text{resp}}$. Plugging in the expression \eqref{psitidalflat} for the tidal field  and using standard Green's function methods,\footnote{It might be easier to solve the equation in Cartesian coordinates by writing the tidal field equivalently  as  $\psi_{\text{tidal}} =\sum_\ell c_{i_1\cdots i_\ell} x^{i_1} \cdots x^{i_\ell}$ \cite{Poisson_Will_2014,Hui:2020xxx}.} one finds the following solution for $\psi_{\text{tidal}}+ \psi_{\text{resp}}$ \cite{Hui:2020xxx,Rai:2024lho}:
\begin{equation}
   \psi_{\text{tidal}}+ \psi_{\text{resp}}= \sum_\ell c_\ell r^\ell P_\ell(\cos\theta) \left[ 1+ \lambda_\ell^{(1)}\frac{(-1)^{\ell+1}}{\Mp^2}\frac{2^{\ell-2}\ell!}{\sqrt{\pi}\,\Gamma(\frac{1}{2}-\ell)}r^{-2\ell-1} \right],
\label{fullpsi1}
\end{equation}
which is valid at linear order in perturbation theory. Since the result \eqref{fullpsi1} has been obtained in the same set of coordinates that we used in \cref{sec:distorted}, we can directly compare \cref{fullpsi1} with \cref{eq:hatpsiphysicalsolution}. Since \cref{eq:hatpsiphysicalsolution} does not contain any inverse falloff in $r$, one thus concludes that $\lambda_\ell^{(1)}=0$ \cite{Rothstein:2014sra,Kol:2011vg,Hui:2020xxx}.

\paragraph{Quadratic response.} {Let us now extend the  matching to quadratic order in perturbation theory. 
This has been previously done for generic cubic operators with arbitrary number of derivatives in \cite{Riva:2023rcm,Iteanu:2024dvx}. In the following, we will  briefly review it in the Weyl  coordinates. We will then discuss how to systematically generalize it to all orders.
}

 The logic proceeds similarly as before. One crucial ingredient, which will  considerably simplify the analysis, is that $\lambda_\ell^{(1)}=0$, which also means that $\psi_{\text{resp}}$ vanishes at linear order.
As a result, this implies that, when we expand the right-hand side of \cref{EinEqsint} to quadratic order, the only terms that contribute are those  quadratic in $E$. This means that  we can still use \cref{Eijexp} at linear order.\footnote{If $\lambda_\ell^{(1)}$ were nonzero, at second order there could be contributions coming from the quadratic expansion of the curvature tensor in the term proportional to  $\lambda_\ell^{(1)}$ on the right-hand side of \cref{EinEqsint}.} Schematically, the equations read 
\begin{equation}
\begin{split}
G_{\mu\nu}^{(2)}
& = - \frac{2}{\Mp^2}\frac{\delta}{\delta g_{\mu\nu}} 
\sum_{\ell=2}^\infty\int \dd\tau  \sum_{\mathclap{\substack{\ell_1, \ell_2 \\ \vert \ell_2- \ell_1 \vert  \leq \ell \leq \ell_1+\ell_2 }}} \lambda_{\ell\ell_1\ell_2}^{(2)}  F(E_{i_L} E_{i_{L_1}}E_{i_{L_2}}) 
\\
& = - 2\Mp^{-2} \sum_{\mathclap{\substack{\ell,\ell_1, \ell_2 \\ \vert \ell_2- \ell_1 \vert  \leq \ell \leq \ell_1+\ell_2 }}} \lambda_{\ell\ell_1\ell_2}^{(2)}  (-1)^\ell
A^{i_L i_{L_1} i_{L_2} }
 \partial_{\langle i_L \rangle}\left(  \delta^{(3)}(\vec{x}) \,
 \partial_{\langle i_{L_1}\rangle}\psi_{\text{tidal}} \partial_{\langle i_{L_2} \rangle}\psi_{\text{tidal}}
  \right) \delta_\mu^0 \delta_\nu^0 ,
\end{split}
\label{EinEqsintQUAD}
\end{equation}
where $A^{i_L i_{L_1} i_{L_2} }$ is a constant tensor, namely a linear combination of all tensor products of $\delta^{ij}$, 
which takes care of all possible independent contractions between the $E$ tensors~\cite{Bern:2020uwk,Riva:2023rcm}, which in \cref{EinEqsintQUAD} we linearized as in \cref{Eijexp}. 
The explicit form of $A^{i_L i_{L_1} i_{L_2} }$ will not be important. The only relevant ingredients are that $A^{i_L i_{L_1} i_{L_2} }$ is coordinate-independent, and that, for fixed number of derivatives per field in a given cubic operator, there is at most one single nontrivial contraction of indices (and, therefore, one single coupling $\lambda_{\ell\ell_1\ell_2}^{(2)}$), compatible with the angular momentum selection rules.

As before, it is convenient to focus on
\begin{equation}
R_{\mu\nu}^{(2)}
 = - 2\Mp^{-2} \sum_{\mathclap{\substack{\ell,\ell_1, \ell_2 \\ \vert \ell_2- \ell_1 \vert  \leq \ell \leq \ell_1+\ell_2 }}} {\lambda_{\ell\ell_1\ell_2}^{(2)}}  (-1)^\ell
A^{i_L i_{L_1} i_{L_2} }
 \partial_{\langle i_L \rangle}\left(  \delta^{(3)}(\vec{x}) \,
 \partial_{\langle i_{L_1} \rangle}\psi_{\text{tidal}} \partial_{\langle i_{L_2} \rangle}\psi_{\text{tidal}}
  \right) \left( \delta_\mu^0 \delta_\nu^0  +\frac{1}{2}\eta_{\mu\nu}  \right) \equiv J_{\mu\nu}^{(2)} .
\label{EinEqsintQUAD2}
\end{equation}
Let us take again the linear combination \eqref{RttRpp}. Keeping only contributions that are at most second order, it  follows from $J^{(2)}_{tt}-J^{(2)}_{\phi\phi}/(r\sin\theta)^2=0$ that it is consistent again to set $\rho=r\sin\theta+ \dots$, up to corrections that are cubic in the field perturbations, or higher. From the $t$--$t$ component of \cref{EinEqsintQUAD2}, one then finds the following inhomogeneous equation for the response field:
\begin{equation}
\vec{\nabla}^2\psi_{\text{resp}} = 2\Mp^{-2} \sum_{\mathclap{\substack{\ell,\ell_1, \ell_2 \\ \vert \ell_2- \ell_1 \vert  \leq \ell \leq \ell_1+\ell_2 }}} {\lambda_{\ell\ell_1\ell_2}^{(2)} } (-1)^\ell
A^{i_L i_{L_1} i_{L_2} }
 \partial_{\langle i_L \rangle}\left(  \delta^{(3)}(\vec{x}) \,
 \partial_{\langle i_{L_1} \rangle}\psi_{\text{tidal}} \partial_{\langle i_{L_2} \rangle}\psi_{\text{tidal}}
  \right)  ,
\label{eqpsiQUAD1}
\end{equation}
where we emphasize that $\psi_{\text{resp}}$ on the left-hand side is a second-order quantity. Note that, since the linear response vanishes and $\psi_{\text{tidal}}$ solves the homogeneous equations of motion, there are no terms quadratic in $\psi$ or of the type $\psi \gamma$ on the left-hand side of \cref{eqpsiQUAD1}.

\Cref{eqpsiQUAD1}  can be easily solved as before. The left-hand side is the usual flat-space Laplacian, while the right-hand side after a Fourier transform is again the product of a fixed number of spatial momenta, up to an irrelevant overall constant tensor, which completely factors out. In the end, one thus finds that $\psi_{\text{resp}} \sim r^{-\ell-1}$, as expected from simple power-counting arguments. Absence of a decaying falloff in the full solution \eqref{eq:hatpsiphysicalsolution} again implies that $\lambda_{\ell\ell_1\ell_2}^{(2)}=0$.

\paragraph{Matching at subleading orders.}
The previous logic can be extended to all subleading orders in perturbation theory in the worldline EFT. The absence of  decaying falloff in the result \eqref{eq:hatpsiphysicalsolution} in the full theory, when compared with the EFT solution,  implies a certain condition on the effective couplings. The particular form of such condition depends in general   on the explicit basis of operators chosen at  given order in the EFT. When there is only one single independent operator at fixed number of derivatives and fields (as at the linear and quadratic orders discussed above), the matching is straightforward, implying the vanishing of the corresponding Love number coupling. The situation can be more involved if more independent operators are present, as matching axisymmetric solutions might not be enough to completely fix all the couplings.
In the latter case, one can however redefine the effective couplings, as prescribed in section~\ref{sec:matchweyl}, in such a way that the matching sets to zero  one of them (corresponding to the only operator that is nontrivial on the axisymmetric  solution). Concretely, one can arrange the EFT in such a way that $\delta \tilde{S}_{\text{int}}/\delta g^{\mu\nu}=0$ in the Einstein equations, when evaluated  on axisymmetric field configurations. 
Therefore, one can just focus at each order on the single operator with coefficient $\lambda_{\ell\ell_1\cdots\ell_n}^{(n)}$ in \cref{Sintalt}, and  repeat iteratively the steps above. 
By induction,  $\lambda_{\ell\ell_1\cdots\ell_n}^{(n)}=0$ at  order $n$  implies  that, at order $n+1$, in $\delta
S_\mathrm{int}/\delta g_{\mu\nu}$ one can replace $E_{ij}$ with the
linear expansion \eqref{Eijexp}. This in turn yields an equation of
the form $R_{\mu\nu}^{(n+1)}\propto (\delta_\mu^0 \delta_\nu^0
+\frac{1}{2}\eta_{\mu\nu})$. After fixing $\rho=r\sin\theta$, one
finds an inhomogeneous equation that contains $\psi$ only, which can
be solved to get $\psi_{\text{resp}}$. The spatial dependence of the
source is localized on the particle's worldline, and is given by a
fixed number of derivatives acting on $\delta^{(3)}(\vec{x})$, in the
rest frame of the particle. The solution's profile is dictated by
power counting, scaling as $\psi_{\text{resp}} \sim r^{-\ell-1}$,
which should be compared with \cref{eq:hatpsiphysicalsolution} to find $\psi_\mathrm{resp}=0$ to all orders in $\hat\psi$.

Let us close with some remarks on what we have accomplished. Focusing on the part of the worldline action containing all the even Love number couplings, we have organized the effective expansion in  a way that, at each order, only one operator is nontrivial  on axisymmetric field configurations. By performing the matching with the Weyl solution, we have shown that all  Love number couplings of such operators vanish for Schwarzschild black holes in general relativity. At the orders where a single independent coupling is present, the matching fixes all the freedom. We stress that, even if the matching was done for axisymmetric solutions, the vanishing of this (infinite) subset of $\lambda$'s holds beyond axisymmetry. The situation is rather similar to other examples of matching in effective field theory. For instance, the coefficient of a given operator in an EFT could be determined by matching with the scattering amplitude of some particular process. Once this is done, that coefficient is fixed, and the same operator can be used to make predictions for the scattering amplitude of other processes.

\section{Symmetries}
\label{sec:sym}

Having established the consequences of the nonlinear Weyl solution at the level of the point-particle EFT, we now turn to study the underlying symmetries.

The keys to the vanishing of an infinite subset of even-parity Love numbers to all orders in perturbation theory shown above
are:
\begin{enumerate}
\item that the fully nonlinear Einstein equations imply a \emph{linear} equation \eqref{psieom} for $\psi$ (which exponentiates to the norm
of the timelike Killing vector);
\item that this equation implies $\hat \psi$ (the perturbation of $\psi$\label{item2}
away from Schwarzschild) develops no tidal tail;
\item that matching with the worldline EFT tells us that the absence of tidal tail for $\hat \psi$
alone is sufficient to guarantee that an infinite subset of tidal response operators vanishes at all orders.
\end{enumerate}
Our task in this section is to dig more deeply into \eqref{item2}: what
symmetries underlie the absence of
a tidal tail for $\hat \psi$?

Let us start by recalling the equation of motion \eqref{psieomSch} for $\hat\psi$ in
Schwarzschild coordinates,
\begin{align}
\label{psieomSch2}
  \partial_r (\Delta \partial_r \hat\psi ) +
  \frac1{\sin\theta}\partial_\theta\left(\sin\theta\partial_\theta
  \hat \psi \right) = 0 \, ,
\end{align}
where $\Delta \equiv r(r -  \rs)$. 
Recall that once we expand $\hat \psi$ in terms of Legendre polynomials of $\cos\theta$, i.e.,~$\hat \psi = \sum_\ell \hat \psi_\ell (r) P_\ell
(\cos\theta)$, the radial modes obey their own Legendre equation (in $\Delta'$),
\begin{align}
\label{psieomSch3}
  \partial_r (\Delta \partial_r \hat\psi_\ell ) - \ell(\ell+1) \hat
  \psi_\ell = 0 \, .
 \end{align}
The radial function $\hat \psi_\ell (r)$
satisfies a linear, second-order differential equation, and is in general a
superposition of two branches of solutions. Their asymptotics
are clear. At large $r$, one branch goes as $r^\ell$ while the other goes as
$1/r^{\ell+1}$. At $r \rightarrow \rs$, one branch goes to a constant while the other
goes as ${\,\rm ln}[(r - \rs)/\rs]$.

The question boils down to this: why is it that once we impose
regularity at the horizon, $\hat \psi_\ell$ at large $r$ is {\it not}
some superposition of both branches $r^\ell$ and $1/r^{\ell+1}$?
Generically, one expects demanding a particular asymptotic behavior
at one end of the spatial region (the horizon) should lead to a
solution that contains both branches of solutions at the other end
(the large-$r$ end). This turns out not to happen here: demanding
regularity at the horizon leads to the absence of $1/r^{\ell+1}$ far
away. We know so from the explicit solution, given in 
\cref{sec:distorted}. Why does it turn out
that way? In other words, our question can be sharply stated as: {\it what is the symmetry story behind the fact
  that the solution for $\hat\psi_\ell$---the one that is regular at the
  horizon---contains no $1/r^{\ell+1}$ tail at
  large $r$?}

Below, in \cref{laddersymmetries}, we present the basic story,
having to do with what are called ladder symmetries \cite{Hui:2021vcv,Berens:2022ebl,Rai:2024lho}.
The key symmetry generator (see \cref{deltaK3} below)
turns out to be part of a bigger algebra, an $\mathfrak{sl}(2,\mathbb
R)$, as explained in \cref{sec:sl2r}.
In \cref{gerochsymmetry}, we discuss the Geroch symmetry that is
well-known to arise in the dimensional reduction process carried out
in \cref{sec:dimred}
\cite{Ehlers:1957zz,Geroch:1970nt,Breitenlohner:1986um,Maison:2000fj,Lu:2007zv,Lu:2007jc},
and how it relates to the ladder symmetries.

\subsection{Ladder symmetries}
\label{laddersymmetries}

We start with the observation that \cref{psieomSch2} is exactly the same as
the equation for a massless, static scalar in Schwarzschild
background, albeit restricted to axisymmetric configurations.
Therefore, we can appeal to the known symmetries discussed in
\cite{Hui:2021vcv,Berens:2022ebl}. Such a system has six symmetries
obeying an  $\mathfrak{so}(3,1)$ algebra, of which three are rotations and three are
generalized conformal transformations. It can be checked that only one
of them maintains the $\phi$ independence of $\hat\psi$, is
non-trivial, and is a symmetry of \cref{psieomSch2}:
\begin{align}
\label{deltaK3}
  \delta_{K_3} \hat\psi = \Delta \cos\theta \partial_r \hat \psi +
\frac12 \Delta' \left( \partial_\theta\sin\theta\,\hat \psi\right)  ,
\end{align}
where $\Delta' = 2r - \rs$, and
$K_3$ refers to the fact that this is a (generalized) special conformal
transformation in the $z$ direction.\footnote{Its connection with
  the standard 
  special conformal transformation can be seen as follows. Take the
  large-$r$ limit, this reduces to $\delta_{K_3} \hat\psi = \left( r^2 \cos\theta \partial_r +
  r \sin\theta \partial_\theta + r \cos\theta \right)
\hat \psi$, which in Cartesian coordinates reads $\delta_{K_3} \hat\psi
 = c_i  (2 x^i \vec x \cdot \vec \partial - \vec x^2 \partial^i +
 x^i)$ for $c_i = (0,0,1)$ in the $z$ direction. 
}
See \cite{Hui:2021vcv,Berens:2022ebl,Hui:2022vbh} for a discussion of how
it arises from a conformal Killing vector of the 3D space
$\dd s^2 = \dd r^2 + \Delta (\dd\theta^2 + \sin^2\theta \dd\phi^2)$ in which
the static, massless scalar effectively resides.

An important observation, following
\cite{Compton:2020cjx,Hui:2021vcv,Berens:2022ebl}, is that\footnote{Useful identities are: 
\begin{align}
\cos\theta  P_{\ell} = \frac{\ell+1}{2\ell+1}  P_{\ell+1} 
+ \frac{\ell}{2\ell + 1} P_{\ell-1} \; , \quad
\sin\theta \, \partial_\theta P_{\ell}
= \frac{\ell (\ell+1)}{2\ell+1} \left( P_{\ell+1} 
- P_{\ell-1} \right)\, .
\end{align}}
\begin{equation}
\label{deltaK3ell}
\delta_{K_3}\left( \hat\psi_\ell P_{\ell}\right)
= - \frac{\ell+1}{2\ell+1} \, D_\ell^+ \hat\psi_\ell  \, P_{\ell+1}
+ \frac{\ell}{2\ell + 1} \, D_\ell^- \hat\psi_\ell \, P_{\ell-1} ,
\end{equation}
with the operators $D^\pm_\ell$  defined by
\begin{align}
\label{Ddef}
D_\ell^+ \equiv -\Delta \partial_r - \frac{\ell+1}{2}\Delta', \qquad
D_\ell^- \equiv \Delta \partial_r - \frac{\ell}{2} \Delta' .
\end{align}
Being a symmetry, $\delta_{K_3}$ maps solutions to solutions.
Thus, if $\hat\psi_\ell P_{\ell}$ is a solution, so is
$\delta_{K_3} (\hat\psi_\ell P_{\ell})$. Looking at the first
term on the right of \cref{deltaK3ell}, since it multiplies
$P_{\ell+1}$,  $D_\ell^+ \hat\psi_\ell$ must correspond to a
radial solution at level $\ell+1$. Similarly, the second term
on the right of \cref{deltaK3ell} multiplies $P_{\ell-1}$,
and so $D_\ell^- \hat\psi_\ell$ must be a radial solution at
level $\ell-1$. Thus, $D_\ell^\pm$ serves to take the level-$\ell$
solution $\hat \psi_\ell$ and raise or lower it to $\hat
\psi_{\ell+1}$ or $\hat \psi_{\ell-1}$, i.e.,~they act as raising and
lowering operators.

Imagine lining up $\hat\psi_\ell P_\ell$ into a giant column vector,
with each element labeled by $\ell$. In group theoretic language,
we would say this forms an infinite representation of $\delta_{K_3}$.
It is non-diagonal, hence the mixing between neighboring $\ell$'s,
which gives rise to the raising and lowering operators.

This representation has a ``ground state,'' the one labeled by $\ell=0$,
for which the equation of motion is very simple:
\begin{equation}
\label{psi0}
\partial_r (\Delta \partial_r \hat \psi_0) = 0 \, .
\end{equation}
A possible ground state is one satisfying $\Delta \partial_r \hat
\psi_0 = D^-_0 \hat \psi_0 = 0$.
This is now a first-order differential equation, for which an issue
associated with a second-order equation does not arise---that is, the
issue of relating two different asymptotics at one end (the horizon)
with two different asymptotics at the other end (the far end).
Indeed, we can see simply that $\hat \psi_0$ equals constant
is a solution. The asymptotic behavior of going to a constant at the
horizon is directly linked to the asymptotic behavior of going as
$r^\ell$ at large $r$ (for $\ell=0$). 

Once this good (regular at the horizon) ground state solution is
identified, one can apply a
string of raising operators to reach the solution at any level $\ell$.\footnote{\label{footnoteD+D-}This is reminiscent of the simple harmonic oscillator, for
  which $(\hat a^\dagger \hat a - n) \Psi_n = 0$. The ground state
  $\Psi_0$ is defined by $\hat a \Psi_0 = 0$ and the excited states
  are reached by acting on $\Psi_0$ repeatedly with the raising
  operator $\hat a^\dagger$. The parallel with what we have here can
  be made explicit by observing that \cref{psieomSch3} can be
  recast as $H_\ell \hat \psi_\ell = 0$, with $H_\ell \equiv
  -\Delta (\partial_r (\Delta \partial_r) - \ell(\ell+1)) =
  D^+_{\ell-1} D^-_\ell - \frac{\ell^2 \rs^2}4$. Note that
  $H_{\ell+1} D^+_\ell = D^+_\ell H_\ell$, $H_{\ell-1} D^-_\ell =
  D^-_\ell H_\ell$, and $D^-_{\ell+1} D^+_\ell - D^+_{\ell-1} D^-_\ell
  = (2\ell + 1)\rs^2/4$. See \cite{Hui:2021vcv} and \cref{sec:ladders}
  for further discussions.
}
And because of the form of $D^+_\ell$, it is plain to see
$\hat \psi_\ell = D^+_{\ell-1} D^+_{\ell-2} \cdots D^+_0 \hat \psi_0$ is a polynomial with
non-negative powers of $r$. Such a level-$\ell$
solution is regular at the horizon and has no $1/r^{\ell+1}$ tidal
tail. Furthermore, an independent solution (of \cref{psieomSch3})
from this one must have
a different asymptotic behavior
at the horizon (the logarithmically divergent one), and can
be discarded, based on the horizon-regularity requirement
discussed in \cref{sec:distorted} and \cref{app:reg}.

In summary, the lack of a $1/r^{\ell+1}$ tidal tail for $\hat \psi$ relies on two
things: (a) the existence of a (generalized) special conformal symmetry
(\cref{deltaK3}), which gives rise to a vertical ladder
structure, allowing one to connect any level $\ell$ solution to the
level $0$ solution, and (b) the good level $0$ solution
obeying a first order differential equation, thus connecting
a single (regular) asymptotic behavior at the horizon with a single
($r^\ell$) asymptotic behavior at large $r$. 

With the EFT matching presented in  \cref{sec:EFT}, the absence of a tidal tail for $\hat
\psi$ in turn implies the vanishing of {a subset of even Love number operators at nonlinear orders}. 
Thus, we can say the vanishing of {such operators} can be traced to the special conformal symmetry
expressed in \cref{deltaK3}, {\it and} the ground state $\ell=0$
solution satisfying a first order equation.

\paragraph{Horizontal ladder symmetries.} The star of the story outlined above
is the (generalized) special conformal symmetry of
\cref{deltaK3}, which is responsible for the ladder structure of
the solutions organized by $\ell$, with operators effecting travels
up and down the ladder. It turns out there are
symmetries that do not involve changing $\ell$, termed horizontal
symmetries \cite{Hui:2021vcv}. For instance, at level $\ell = 0$, it is obvious
$\delta \hat\psi_0 = \Delta \partial_r\hat \psi_0$ is a symmetry, i.e.,~$\delta\hat\psi_0$ is a solution if $\hat\psi_0$ is a solution of
\cref{psi0}. Calling $Q_0 \equiv \Delta \partial_r$, it
can be shown (and is intuitive) that $Q_\ell$, defined as
$Q_\ell \equiv     D^+_{\ell-1} D^+_{\ell-2} \cdots D^+_0  Q_0 D^-_1
\cdots D^-_{\ell-1} D^-_\ell$, generates a symmetry at level $\ell$, i.e.,~that
$Q_\ell \hat\psi_\ell$ is a solution if $\hat \psi_\ell$ is a
solution. The implied conserved charge at each $\ell$---conserved in the sense of
being $r$ independent---can be used to connect asymptotic behavior
at the horizon with asymptotic behavior at large $r$, thus offering
another way to understand the phenomenon of no tidal tail.
Details can be found in \cite{Hui:2021vcv,Berens:2022ebl}.
An additional interesting observation: the conserved charge mentioned above
turns out to be the Wronskian squared, where the Wronskian is that
between the $\hat \psi_\ell$ of interest and the horizon-regular solution
\cite{BenAchour:2022uqo}.\footnote{\label{noteWron}By the Wronskian, we mean $W[\hat \psi_\ell {}^{\rm reg},
  \hat\psi_\ell ] = \hat
  \psi_\ell {}^{\rm reg} \Delta \partial_r \hat\psi_\ell - \hat
  \psi_\ell \Delta \partial_r \hat\psi_\ell {}^{\rm reg}$, where
  $\hat \psi_\ell^{\rm reg}$ is the regular solution
  and $\hat \psi_\ell$ is the field configuration
  of interest.
  }
  This is further discussed in \cref{sec:ladders}.

The term {\bf ladder symmetries} refers to both the (generalized)
special conformal symmetry \eqref{deltaK3}, which gives rise
to the vertical ladder structure, and the horizontal symmetries, one
for each multipole $\ell$.

\subsection{\texorpdfstring{SL$(2,\mathbb R)$ symmetry}{SL(2,R)
    symmetry}}
\label{sec:sl2r}
It is worth asking:
is it possible \cref{psieomSch2} contains more (first order) symmetries than the
$\delta_{K_3}$ identified in \cref{deltaK3}? To answer this it is helpful to go back to the original form
of the equation in $(\rho, z)$ coordinates:
\begin{equation}\label{eq:psi-eom2}
\left(\partial_\rho^2+\frac1\rho\partial_\rho+\partial_z^2\right) \hat
\psi=0
\, .
\end{equation}
This describes a free, massless scalar living in a fictitious, 3D flat
space, in cylindrical coordinates.
We know such a scalar in principle has conformal invariance with 10 independent
symmetry generators. But to keep $\hat\psi$ independent of the azimuthal
angle $\phi$, only three out of the ten survive: translations in
the $z$ direction $P$, dilations $D$, and special conformal transformations
in the $z$ direction $K$ \cite{Miller_1984}:
\begin{equation}
P = \partial_z,\quad D =
-\left(\frac12+\rho\partial_\rho+z\partial_z\right),\quad K = 2\rho
z\partial_\rho+\left(z^2-\rho^2\right)\partial_z+z \, .
\end{equation}
It is straightforward to check they form an $\mathfrak{sl}(2,\mathbb R)$ algebra,
\begin{equation}
[D,P]=P, \quad [D,K] = -K, \quad [P,K]=-2D.
\label{sl2Ralgebra}
\end{equation}
In Schwarzschild coordinates, i.e.,~with $\rho = \Delta \sin\theta$ and $z = (\Delta' \cos\theta )/ 2$, these operators
take the form:
\begin{subequations}
\label{PDKSch}
\begin{align}
P &= {\rm e}^{2\gamma_\sch}\left(\cos\theta\partial_r-\frac{\Delta'}{2\Delta}\sin\theta\partial_\theta\right),\\
D &= {\rm e}^{2\gamma_\sch}\left(-\frac12\Delta'\partial_r + \frac{\rs^2}{4\Delta}\sin\theta\cos\theta\partial_\theta  \right)- \frac12,\\
K &= \Delta\cos\theta\partial_r +
    \frac{\Delta'}2\left(\sin\theta\partial_\theta+\cos\theta\right)
    +\frac{\rs^2}4{\rm e}^{2\gamma_\sch}\left(\cos\theta\partial_r-\frac{\Delta'}{2\Delta}\sin\theta\partial_\theta\right)
    \, ,
\end{align}
\end{subequations}
where we remind the reader that we have defined $\gamma_\sch$ in \cref{gammaSch}.
We recognize the combination
\begin{align}
\left(K-\frac{\rs^2}4P\right)\hat\psi =
  \Delta\cos\theta\partial_r\hat\psi + \frac{\Delta'}2 \sin\theta
  \partial_\theta \hat\psi + \frac{\Delta'}2 \cos\theta \hat\psi
\end{align}
as precisely $\delta_{K_3} \hat\psi$ given earlier in \cref{deltaK3}.\footnote{It is worth commenting on the difference between $K$ and
  $K_3$. The transformation effected by $K$ is the standard special conformal transformation in
  the $z$ direction, in a fictitious 3D flat space, $\dd s^2 = \dd\rho^2 +
  \dd z^2 + \rho^2 \dd\phi^2$. The transformation effected by $K_3$, defined in
  \cref{deltaK3}, is a (generalized) special conformal
  transformation in the $z$ direction, in the space $\dd s^2 = \dd r^2 + \Delta \dd \theta^2 + \Delta
  \sin^2\theta \dd\phi^2$, which is the space effectively seen by a
static massless scalar on Schwarzschild background \cite{Hui:2021vcv}, as well as the Einstein-frame metric in the dimensional reduction performed in \cref{sec:dimred}. The two coincide
in the flat space limit, $\rs \rightarrow 0$.
}
In summary, the full symmetry algebra for $\hat \psi$ is
$\mathfrak{sl}(2,\mathbb R)$, of which the combination $\delta_{K_3} \hat\psi$ gives the
simplest route to the ladder structure and its implied absence of tidal
tail.

Let us close with an observation that echoes one made earlier
towards the end of \cref{sec:distorted}. There is nothing sacred
about the $(r,\theta)$ Schwarzschild coordinates, even though it is a natural
choice for thinking about the Schwarzschild background. 
Starting from the $(\rho, z)$ Weyl coordinates, we could choose
$({\mathcal R}, \vartheta)$ coordinates, with $\rho = \mathcal R \sin
\theta$ and $z = \mathcal R \cos\theta$, which are in a sense better suited to the fictitious flat space that $\hat\psi$ effectively
lives in. The $D$, $P$, $K$ operators take the same
form as in \cref{PDKSch}, with the replacement $r \rightarrow
\mathcal R$, $\theta \rightarrow \vartheta$ and $\rs \rightarrow 0$.
The ladder structure follows from the actions of $K$ and $P$, and one
can run essentially the same symmetry argument as before.\footnote{
  That the radial solutions in this case are so
  simple---$\mathcal R^\ell$ and $1/\mathcal R^{\ell+1}$---might seem to make a symmetry
  argument an overkill. Nonetheless, the
  symmetry argument is what connects asymptotic behavior
  at one end with asymptotic behavior at the other. The $(\mathcal R,
  \vartheta)$ coordinates are special in that, for each branch of
  solution, the asymptotic behavior far away is the {\it same} as the
  asymptotic behavior closeby.}
This way of proceeding is simpler,
because $K$ and $P$ expressed in $(\mathcal R, \vartheta)$ coordinates takes
a simpler form, though at the cost of a more complicated expression for
the Schwarzschild background.\footnote{One might also wonder: how about studying the
  tidal response in the original Weyl coordinates $(\rho, z)$?
  In that case, the general separable solution that is regular on the symmetry axis is
  $(a \mathrm e^{\lambda z}+b\mathrm{e}^{\lambda z})J_0(\lambda\rho)$,
  where $\lambda$ is a separation constant, $J_0$ is the Bessel
  function, and $a$ and $b$ are constant coefficients. In this case, it is more
  cumbersome to investigate the large-distance limit, and draw
  conclusions about the tidal tail or lack thereof.}

\subsection{Geroch symmetry}
\label{gerochsymmetry}

The dynamics of the even-parity, nonlinear black hole distortions discussed in this work arise from dimensionally reducing four-dimensional general relativity along the Killing vectors $\xi$ and $\eta$. It is well-known that this dimensional reduction leads first to the $\mathrm{SL}(2,\mathbb R)$ Ehlers group in three dimensions, and then to the infinite-dimensional Geroch group in two dimensions \cite{Geroch:1970nt,Breitenlohner:1986um,Maison:2000fj,Lu:2007zv,Lu:2007jc}. Initially discovered as a solution-generating technique \cite{Geroch:1970nt,Stephani:2003tm}, the Geroch symmetry is remarkably powerful; from a Minkowski space seed, one can generate complicated solutions such as Kerr-NUT using a Geroch transformation \cite{Maison:2000fj,Katsimpouri:2012ky}. But with great power comes great responsibility, and for the Geroch group the responsibility is usually to solve a rather difficult inverse-scattering or Riemann--Hilbert problem \cite{Breitenlohner:1986um,Maison:2000fj,Katsimpouri:2012ky,Katsimpouri:2015nqc}. Fortunately our responsibility is rather less heavy, as we need only the \emph{infinitesimal} version of the Geroch transformation \cite{Lu:2007zv,Lu:2007jc}\footnote{Note that \cite{Lu:2007zv,Lu:2007jc} worked with a dimensional reduction along two spacelike directions, so some expressions in this section will differ from those references accordingly. See also \cite{Cardoso:2017cgi} for deformations of the Schwarzschild metric obtained via the inverse scattering technique.} to study the Love numbers. The problem is simplified even more by our assumption that the spacetime is static (i.e., that the $\chi$ field introduced in \cref{app:dimred} is turned off).

We discuss the action of the Geroch symmetry on $\hat\psi$ in \cref{app:geroch}. For the reader in a hurry we restrict ourselves here to a summary of the points that are important for our purposes. The infinite-dimensional nature of the Geroch symmetry is encoded by a constant spectral parameter $w$, in terms of which the infinitesimal action on $\hat\psi$ is \cite{Lu:2007jc}
\begin{equation}
\delta \hat\psi =\frac{w}{\sqrt{\rho^2+(w-z)^2}}. \label{eq:perry-sym-static}
\end{equation}
The function on the right-hand side has some interesting properties. First, it is a solution of the Laplace equation \eqref{eq:psi-eom} for any value of $w$. To see what kind of solution it is, we expand it at large $w$, finding that it takes a rather simple form in the polar coordinates $(\rho,z)=(\mathcal R \sin\vartheta,\mathcal R\cos\vartheta)$,
\begin{equation}
\frac{w}{\sqrt{\rho^2+(w-z)^2}} = \displaystyle\sum_{n\geq0}w^{-n}\mathcal R^nP_n(\cos\vartheta).
\end{equation}
Expanding the $\delta$ operator in \cref{eq:perry-sym-static} as $\delta=\sum w^{-n}\delta_{(n)}$, we see that each of the symmetry operators $\delta_{(n)}$ shifts $\hat\psi$ by the pure growing solution at level $n$ in the separable solution in polar coordinates,
\begin{equation}\label{eq:delta-n}
\delta_{(n)}\hat\psi = \mathcal R^n P_n (\cos\vartheta).
\end{equation}
Acting on a separable mode of the form $\hat\psi=\hat\psi_k(\mathcal R)P_k(\cos\vartheta)$, this becomes the radial symmetry
\begin{equation}
\delta_{(n)}\hat\psi_k = \mathcal R^n \delta_{kn}.
\end{equation}
The conserved quantity associated to this symmetry is again a Wronskian,\footnote{If we work with the symmetry operator \eqref{eq:delta-n} then the conserved current satisfying $\partial_a J^a=0$ is a gradient Wronskian,
\begin{equation}
J_a = \rho\left(\hat\psi\partial_a\delta\hat\psi - \delta\hat\psi\partial_a\hat\psi\right).
\end{equation}
Expanding in powers of $1/w$ we find the level-$n$ conserved current,
\begin{equation}
J ^{(n)}_a = \rho \left[\mathcal{R}^n P_n(\cos\vartheta)\partial_a\hat\psi - \hat\psi\partial_a\left(\mathcal{R}^n P_n(\cos\vartheta)\right)\right] .
\end{equation}}
\begin{equation}
\tilde Q_n \equiv \mathcal R^2\left(\mathcal R^n\partial_{\mathcal R}\hat\psi-\hat\psi\partial_{\mathcal R}\mathcal R^n\right).
\end{equation}
As discussed in \cite{BenAchour:2022uqo} and in \cref{sec:ladders}, the shift-by-a-solution symmetry has as its conserved charge the Wronskian with the reference solution, while the associated linear symmetry, which may be obtained via a Poisson bracket, has as its conserved quantity the square of the Wronskian. While the Geroch symmetry \eqref{eq:perry-sym-static} is most conveniently expressed in polar coordinates, an $\ell$ mode in Schwarzschild coordinates is just a linear combination of $n$ modes in polar coordinates, so we may identify the horizontal symmetry at a given $\ell$ with a linear combination of $\delta_{(n)}$.

\section{Discussion}
\label{sec:discuss}

In this work, we studied nonlinear tidal effects of Schwarzschild black holes in 4D general relativity.
We focused on the Weyl metric \eqref{eq:weyl}, which provides the most general framework for static and axisymmetric spacetimes, and studied the solution to the full nonlinear Einstein equations.
Using the framework  of the point-particle EFT, we restricted ourselves to operators involving the electric component of the curvature tensor only, and introduced a set of nonlinear Love number couplings which capture the (conservative) nonlinear tidal response induced by static parity-even perturbations.
By performing the matching with the full solution, we showed that {an infinite subset of such Love numbers vanish to all orders in perturbation theory}. 

In the second part of the work, we proposed a fully nonlinear symmetry explanation for the vanishing of the Love numbers. We showed that there exists in 4D general relativity a hidden structure of ladder symmetries which is responsible for the absence of a tidal tail in the nonlinear solution \eqref{eq:hatpsiphysicalsolution}. The ladder generator belongs to an $\mathfrak{sl}(2,\mathbb R)$ algebra (see \cref{sl2Ralgebra}), whose action is a symmetry of the perturbation equation \eqref{psieomSch}.
The symmetries, which act fully nonlinearly at the level of the metric perturbations, recover the ones introduced in \cite{Hui:2021vcv} in the linear regime, when written in the same  coordinates. 
In addition to the ladder generator, we  showed that there are ``horizontal'' types of symmetries, which do not mix different $\ell$'s in harmonic space, that are connected to the Geroch group, resulting from  dimensional reduction of general relativity from  four to two dimensions.

There are several interesting directions that we envision and  leave for future investigation. 
{First of all, it would be particularly interesting to relax the assumption of axisymmetry and show that all the remaining couplings, which we were not able to fix, are also zero for black holes in general relativity.
In addition,} our analysis here was confined to parity-even perturbations only. However, explicit calculations  have shown that both even and odd static second-order perturbations  display a very simple structure,  generically taking the form of finite polynomials, irrespective of their magnetic quantum numbers \cite{Riva:2023rcm,Iteanu:2024dvx,Poisson:2009qj}. It is therefore tempting to speculate that it might be possible  to find a more convenient description of perturbation theory that allows one to resum nonlinearities and extend the ladder symmetries to nonlinear order including in the odd sector.
In this work, we primarily focused on the solution and symmetries of $\hat{\psi}$.
This was enough for our purposes, because that  is the only ingredient
that we needed to compute the Love numbers and perform the matching
with the worldline EFT. However, at the level of the infrared effective
action, there ought to exist a formulation of symmetries acting on the
whole metric, applicable even away from axisymmetry.
Finally, it would be interesting to go beyond the static assumption and consider the more general case of rotating black holes, as well as to study the case of charged solutions, and higher dimensional spacetimes. We leave these and related questions for  future work.

\paragraph{Acknowledgements.}
We would like to thank Roman Berens, Austin Joyce, Don Marolf, Daniel McLoughlin,
Riccardo Penco, Malcolm Perry, Massimiliano Maria
Riva, Mar\'ia Rodr\'iguez, Nikola Savi\'c, John Staunton, and Filippo
Vernizzi for useful discussions.
{We are thankful to Nikola Savi\'c and Filippo Vernizzi for pointing out a correction
in section~\ref{sec:EFT}.}
OCS is supported by a PhD Joint Program of the International Research Center for Fundamental Scientific Discovery (IRC Discovery).
LH acknowledges support from the DOE DE-SC011941 and a Simons
Fellowship in Theoretical Physics.
The research of LS has been funded, in part, by the French National Research Agency (ANR) under project ANR-24-CE31-1097-01, and by the Programme National GRAM of CNRS/INSU with INP and IN2P3 co-funded by CNES.
ARS's research was partially supported by funds from the Natural Sciences and Engineering Research Council (NSERC) of Canada. Research at the Perimeter Institute is supported in part by the Government of Canada through NSERC and by the Province of Ontario through MRI.
LH and SW are grateful to the Universit\'e Paris Cit\'e and the
Astroparticule et Cosmologie (APC) laboratory for hospitality and support during a
research visit. 

\appendix

\section{Dimensional reduction: peeling the general relativistic onion}
\label{app:dimred}

In \cref{sec:dimred} we reduced the dynamics of general relativity from 4D to 2D along isometry directions. With every dimensional reduction, some of the metric components become lower-spin fields. When we reduce all the way down to 2D, all that remains of the metric is the conformally flat $g_{2,ab}$, leaving us in effect with just a theory of scalars. In this way we peel layer by layer off of the complex dynamics of general relativity until we are left with something much more tractable (and indeed integrable \cite{Maison:1978es}).

For concreteness we first reduce along $t$ and then along $\phi$. For the first dimensional reduction, we decompose $x^\mu=(t,x^i)$ and write the metric as
\begin{equation}
\dd s_4^2 = -\mathrm{e}^{-\psi}(\dd t+A)^2+\mathrm{e}^\psi\dd s_3^2.
\end{equation}
This is a valid ansatz for any four-dimensional Lorentzian spacetime, but our assumption that $\partial_t$ is Killing restricts the three-dimensional fields $(\psi,A,g_3)$ to be independent of $t$. The 4D Einstein--Hilbert action decomposes into
\begin{equation}
\sqrt{-g_4}R_4 = \sqrt{g_3}\left(R_3-\frac12(\partial\psi)^2+\frac14\mathrm{e}^{-2\psi}(\dd A)^2\right),
\end{equation}
where we have dropped total derivatives. We see that $g_3$ is the Einstein-frame metric, as it is minimally coupled to $\psi$. This is the reason for choosing the factor $\mathrm{e}^\psi$ in front of $\dd s_3^2$, which as usual one may do by virtue of a conformal transformation.

The vector field $A$ has a Maxwell term non-minimally coupled to the dilaton $\psi$. In vacuum its equation of motion is
\begin{equation}
\dd(\mathrm{e}^{-2\psi}{\star_3\dd A}) = 0 = \nabla^i\left(\mathrm{e}^{-2\psi}F_\Ij\right),
\end{equation}
where $\star_3$ is the Hodge star associated to the 3-metric $g_3$. This implies the existence of a scalar potential $\chi$ dual to $A$,\begin{equation}
\star_3\dd A = \mathrm{e}^{2\psi}\dd\chi.\label{eq:dual-A}
\end{equation}
One can exchange $A$ for $\chi$ at the level of the action by introducing $\chi$ as an auxiliary field (via a perfect square so as not to affect the dynamics) and integrating out $A$. The effect is the same as simply imposing \cref{eq:dual-A} in the action,
\begin{equation}
\sqrt{-g_4}R_4 = \sqrt{g_3}\left(R_3-\frac12(\partial\psi)^2-\frac12\mathrm{e}^{2\psi}(\partial\chi)^2\right).\label{eq:EH-3D}
\end{equation}
The scalars $(\psi,\chi)$ form an $\mathrm{SL}(2,\mathbb R)/\mathrm{SO}(2)$ sigma model. Note the role played by the reduction to $D=3$ in particular. It is only in 3D that a massless vector is dual to a scalar; in general it is dual to a $(D-3)$-form and does not form a sigma model with $\psi$.

The dimensional reduction along the $\phi$ direction proceeds along largely similar lines, with one exception: because of conformal invariance in 2D we must treat the $g_{\phi\phi}$ metric component and the conformal factor in front of the 2-metric as independent,
\begin{equation}\label{eq:g3}
\dd s^2_3 = \rho^2\left(\dd \phi + b\right)^2 + \mathrm{e}^{2\gamma}\dd s_2^2.
\end{equation}
As a result there is an unavoidable conformal factor in front of the 2D Einstein--Hilbert term,
\begin{equation}
\sqrt{-g_4}R_4 = \sqrt{g_2}\rho\left[R_2 - \frac14\rho^2\mathrm{e}^{-\gamma}(\dd b)^2+\frac{2}{\rho}\partial\rho\cdot\partial\gamma -\frac12(\partial\psi)^2-\frac12\mathrm{e}^{2\psi}(\partial\chi)^2 \right].\label{eq:L-2D-full}
\end{equation}

None of the fields $(\rho,\gamma,b,g_2)$ is dynamical, and the two gravitational degrees of freedom can be taken to live in the $\mathrm{SL}(2,\mathbb R)$ fields $(\psi,\chi)$. For the 2-metric, because every 2D pseudo-Riemannian manifold is locally conformally flat, in a suitable coordinate system we can set $g_{2,\Ij}=\delta_\Ij$ by absorbing the conformal factor into $\gamma$. It is also straightforward to set $b=0$ by integrating its equation of motion,\footnote{Herein $\star\equiv\star_2$ is the 2D Hodge star.}
\begin{equation}
\dd\left(\rho^3\mathrm{e}^{-\gamma}\star\dd b\right)=0\quad \Longrightarrow \quad\rho^3\mathrm{e}^{-\gamma}\star\dd b=\mathrm{const.}
\end{equation}
and invoking boundary conditions, such as asymptotic flatness, to set the constant to zero. Varying the action with respect to $\gamma$ we find, in vacuum,
\begin{equation}
\Box_2\rho = \nabla^2\rho = 0,
\label{eq:rrho}
\end{equation}
where $\Box_2=g_2^\Ij\nabla_{i}\nabla_{j}$ and $\nabla^2=\delta^\Ij\partial_i\partial_j$ in coordinates where $g_2\propto\delta$.
The equality between the d'Alembertian and the flat-space Laplacian in 2D is a consequence of the conformal invariance of $\sqrt{g_2}g_2^\Ij$.
Because any solution for $\rho$ is a harmonic function on $\mathbb{R}^2$, as long as $\dd\rho\neq0$ we may think of it as a harmonic coordinate rather than a field. The natural choice for the second harmonic coordinate is its dual scalar $z$, defined by
\begin{equation}
\dd z = -\star\dd\rho.\label{eq:z-form}
\end{equation}
The coordinates $x^a=(\rho,z)$ are commonly known as \emph{Weyl canonical coordinates}.

We are left with three fields to solve for to fully reconstruct the metric, namely the $\mathrm{SL}(2,\mathbb R)$ coset fields $(\psi,\chi)$ and the 2D conformal factor $\gamma$. The latter may be found from the former via the 2D Einstein equation, which in Weyl canonical coordinates is
\begin{equation}\label{eq:lambda0}
\partial_{(i}\rho\partial_{j)}\gamma - \frac12(\partial\rho\cdot\partial\gamma) \delta_\Ij= \frac14\rho\left[\partial_i\psi\partial_j\psi-\frac12(\partial\psi)^2\delta_\Ij + \mathrm{e}^{2\psi}\left(\partial_i\partial_j\chi-\frac12(\partial\chi)^2\delta_\Ij\right) \right].
\end{equation}
There is only one independent component, which we can isolate by projecting along $\partial_i\rho$,\footnote{We may also write this in forms notation as
\begin{equation}
\dd\gamma = \frac14\rho\left[\partial_\rho\psi\dd\psi - \partial_z\psi\star\dd\psi + \mathrm{e}^{2\psi}\left(\partial_\rho\chi\dd\chi - \partial_z\chi\star\dd\chi\right)\right].
\end{equation}}
\begin{equation} \label{eq:lambda}
\partial_i\gamma = \frac12\rho\left(\partial_{\langle i}\psi\partial_{j\rangle}\psi+\mathrm{e}^{2\psi}\partial_{\langle i}\chi\partial_{j\rangle}\chi\right)\partial^j\rho,
\end{equation}
where angular brackets denote traceless symmetrization. Given a solution for $(\psi,\chi)$, one can obtain $\gamma$ by line integration. The equation of motion for the $\mathrm{SL}(2,\mathbb R)$ fields is conventionally expressed in terms of the Ernst potential,
\begin{equation}
\mathcal{E} \equiv \mathrm{e}^{-\psi} + i\chi,
\end{equation}
where it goes by the name of the Ernst equation,
\begin{equation}
\dd(\rho\star\dd\mathcal E)=\frac{\rho}{\operatorname{Re}(\mathcal E)}\dd\mathcal E\wedge\star\dd\mathcal E.
\end{equation}
For spacetimes that are static rather than stationary, there are no time-space cross terms in the metric, so $A_i$ and therefore $\chi$ vanish. Here something truly remarkable occurs: the equation of motion for $\psi$, the only dynamical field in the picture, becomes \emph{linear}, despite the fact that we are working with fully nonlinear general relativity,
\begin{equation}
\dd(\rho\star\dd\psi)=0. \label{eq:eom-psi_forms}
\end{equation}
In Weyl canonical coordinates this takes a particularly suggestive form,
\begin{equation}
\left(\partial_\rho^2+\frac1\rho\partial_\rho+\partial_z^2\right)\psi=0.
\end{equation}
We may interpret this as the Laplace equation in a fictitious three-dimensional flat space with cylindrical coordinates $(\rho,\phi,z)$, restricted to a slice of constant $\phi$.

This linearity property is the principal reason we focus in this work on static rather than stationary spacetimes. Invariance under $t\to-t$ restricts us to even-parity distortions of non-rotating black holes. Inclusion of the $t$--$i$ cross terms by keeping $\chi\neq0$ would allow us to consider two additional cases of physical relevance, namely $\omega=0$ odd-parity perturbations of Schwarzschild black holes and $\omega=m=0$ perturbations of Kerr black holes. 
Even though the Ernst equation is nonlinear, the theory remains integrable \cite{Maison:1978es}, or, equivalently, invariant under the infinite-dimensional Geroch symmetry \cite{Geroch:1970nt,Breitenlohner:1986um,Maison:2000fj,Lu:2007jc}. We consider this in future work.

At the end of the day we are left with the famous \emph{Weyl metric},
\begin{equation}
\dd s^2_4 = -\mathrm{e}^{-\psi}\dd t^2 + \mathrm{e}^\psi\left[\mathrm{e}^{2\gamma}(\dd\rho^2+\dd z^2)+\rho^2\dd\phi^2\right],
\end{equation}
which is the general ansatz for a static and axisymmetric vacuum spacetime in general relativity \cite{Weyl:1917gp,Stephani:2003tm}. 

\section{Regularity condition for the distorted potential \texorpdfstring{$\hat{\psi}$}{}}
\label{app:reg}

In this appendix we summarize the proof that, for a distorted black hole described by $\psi = \psi_\sch + \hat{\psi}$, $\hat{\psi}$ must be regular at the horizon, following from the regularity of the scalars $\xi_{\mu}\xi^{\mu}$, $\eta_{\mu}\eta^{\mu}$, and $\nabla_{\mu}\xi_{\nu}\nabla^{\mu}\xi^{\nu}$ \cite{Geroch:1982bv}.

We first argue that the horizon of a distorted black hole must also be located at $\rho=0$ along the $z$ axis. Since, at the horizon, the norm of the timelike Killing vector, $\xi_{\mu}\xi^{\mu} = -\mathrm{e}^{-\psi} $, has to vanish, and the norm of the axial Killing vector, $\eta_{\mu}\eta^{\mu} =\rho^2 \mathrm{e}^\psi$, remains finite, their product $\xi_{\mu}\xi^{\mu}\eta_{\nu}\eta^{\nu} =\rho^2$ must vanish on the horizon. Therefore, the horizon can only be located at $\rho=0$. It follows that the Einstein equation gives the following equation for $\psi$,\footnote{Despite the delta-function source, the Ricci tensor, in particular the component $R_{tt} =-\frac{1}{2}\mathrm{e}^{-2(\psi + \gamma)} \nabla^2 \psi$, is still vanishing at the horizon.}
\begin{align} \label{eqn:laplacesource}
  \nabla^2 \psi = -\frac{2}{\rho}\delta(\rho) \lambda(z) + S_{\rm outside}(\rho,z)  ,
\end{align}
 where the source at the horizon, $\frac{1}{\rho}\delta(\rho) \lambda(z)$,  generates the irregular part of $\psi$ around the horizon, and $S_{\rm outside}(\rho,z)$ denotes any sources that are located away from the horizon.
 
While the solution $\psi$ can be obtained using the Green's function $-\frac{1}{4\pi \sqrt{\rho^2 + (z-w)^2}}$, instead of looking at the full solution, we only need the $\rho \to 0 $ behavior of $\psi$. Using Gauss's law around an extremely small tube around the $z$ axis, one concludes that 
\begin{align} \label{eqn:psiasymp}
\psi \to-2\lambda(z) \ln \rho + C_1(z)  \quad \mbox{ for }  \rho \to 0,
\end{align}
where $C_1(z)$ is some bounded function at the horizon.  Since the norms of the Killing vectors, $-\mathrm{e}^{-\psi}$ and $\rho^2 \mathrm{e}^\psi$, are bounded at the horizon, one can immediately conclude that 
\begin{align}
      -c  \le \psi \le -2\ln \rho +c
\end{align}
for some positive constant $c$. In order for \cref{eqn:psiasymp} to satisfy the bound, we require
\begin{align}
0 \le \lambda(z) \le1.
\end{align}
Now consider the scalar\footnote{The regularity of a solution is usually evaluated using the Kretschmann scalar $R_{\mn\ab}R^{\mn\ab}$, which is the simplest non-vanishing scalar one can build out of the curvature of a Ricci-flat spacetime. The existence of Killing vectors allows us to construct simpler curvature-like scalars; a notable practical distinction is that the Kretschmann scalar has a rather more complicated dependence on $\gamma$.}
\begin{align}
  \nabla_{\mu}\xi_{\nu}\nabla^{\mu}\xi^{\nu} = -\frac{1}{2} \mathrm{e}^{-2(\psi+ \gamma)} \left[ (\partial_\rho\psi)^2 + (\partial_z\psi)^2  \right].
\end{align}
At the horizon, this quantity has to be finite. From
\cref{eq:lambda-main,eqn:psiasymp} we have $\gamma \to \lambda(z)^2
\ln \rho+C_2(z)$, where $C_2(z)$ is bounded at the horizon.
Therefore,
\begin{align}
\nabla_{\mu}\xi_{\nu}\nabla^{\mu}\xi^{\nu} \overset{\rho\to0}{\to} -2 \rho^{-2 (\lambda(z)-1)^2 } \left[ \lambda(z)^2 + \rho^2 (C_1'(z) - \lambda'(z) \ln \rho)^2 \right]\mathrm{e}^{-2C_1(z)-2C_2(z)} .
\end{align}
At the horizon, $\lambda'(z)$ , $C_1(z)$ and $C_1'(z)$ should be finite. Finiteness of $\nabla_\mu\xi_\nu\nabla^\mu\xi^\nu$ at $\rho \to 0$ leads us to conclude that $\lambda(z) =1$ almost everywhere at the horizon. Therefore, the problem~\eqref{eqn:laplacesource} reduces to
\begin{align}
  \nabla^2 \psi = -\frac{2}{\rho}\delta(\rho)\theta(GM-z)\theta(z+GM)+ S_{\rm outside}(\rho,z) ,
\end{align}
where we have set the location of the horizon to be within the range $z \in [-GM,GM]$ without loss of generality. We know that the first part,  $-\frac{2}{\rho}\delta(\rho)\theta(GM-z)\theta(z+GM)$, gives us precisely the ``background" potential $\psi_\sch$ of a Schwarzschild black hole of mass $M$. Therefore $\hat{\psi}$ should satisfy $\nabla^2 \hat{\psi} =0$ {\it exactly} in the vicinity of the horizon. From the properties of the Laplace equation we conclude that $\hat{\psi}$ must be finite and regular at the horizon.

\section{Black hole perturbation theory}
\label{app:BHPT}

If we consider the tidal distortion to be a small perturbation of the background, $\hat\psi\ll\psi_\sch$, we obtain a theory of the perturbative static response. This must of course be equivalent to the standard treatment within black hole perturbation theory (BHPT) \cite{Regge:1957td}, though the dictionary between the formalisms turns out to be slightly non-trivial, as we detail in this appendix.

The Weyl metric \eqref{eq:weyl} for the distorted black hole is \cite{Geroch:1982bv}
\begin{equation}\label{eq:met-GH}
\dd s^2 = -\mathrm{e}^{-\hat\psi}f(r)\dd t^2+\mathrm{e}^{\hat\psi}\left[\mathrm{e}^{2\hat\gamma}\left(\frac1{f(r)}\dd r^2+r^2\dd\theta^2\right) + r^2 \sin^2\theta\dd \phi^2\right],
\end{equation}
where the coordinates $(r,\theta)$ are related to the Weyl canonical coordinates $(\rho,z)$ by
\begin{equation}\label{eq:sch-weyl-coord}
\rho = \sqrt{r(r-\rs)}\sin\theta,\qquad z = \left(r-\frac\rs2\right)\cos\theta.
\end{equation}
Linearizing in the distortion fields, the metric \eqref{eq:met-GH} is
\begin{equation}\label{eq:met-GH-lin}
\dd s^2 = -(1-\hat\psi)f(r)\dd t^2+\left[(1+\hat\psi+2\hat\gamma)\left(\frac1{f(r)}\dd r^2+r^2\dd\theta^2\right) + r^2 \sin^2\theta(1+\hat\psi)\dd \phi^2\right] + \mathcal{O}(\hat\psi^2,\hat\gamma^2,\hat\psi\hat\gamma).
\end{equation}

Our Schwarzschild coordinates $(r,\theta)$ are defined covariantly as functions of $\rho$ and $z$. They are not, however, the same as the Schwarzchild coordinates conventionally used in BHPT in Regge--Wheeler gauge, which we label $(\tilde r,\tilde\theta)$. The linearized metric for even-parity perturbations is\footnote{We have used two of the non-dynamical Einstein equations --- $H_1=0$ and $H_2=H_0$, in standard notation --- to write the metric in this form.}
\begin{equation}
\dd s^2 =  -f(\tilde r)(1-H_0)\dd t^2 + \frac{1+H_0}{f(\tilde r)}\dd \tilde r^2 + \tilde r^2(1+\mathcal K)\left(\dd\tilde\theta^2+\sin^2\tilde\theta\dd\phi^2\right),\label{eq:RW-metric}
\end{equation}
where $H_0$ and $\mathcal K$ are shorthand for the mode sums
\begin{subequations}
\begin{align}
H_0 &\equiv \displaystyle\sum_\ell H_{0,\ell}(\tilde r)Y_{\ell}(\tilde\theta),\\
\mathcal K&\equiv \displaystyle\sum_\ell \mathcal K_{\ell}(\tilde r)Y_{\ell}(\tilde\theta).
\end{align}
\end{subequations}
The $m=0$ spherical harmonics $Y_\ell(\theta)\equiv Y_{\ell,0}(\theta)$ are related to the Legendre polynomials $P_\ell(\cos\theta)$ by a normalization factor,
\begin{equation}
Y_\ell(\theta) = \sqrt{\frac{2\ell+1}{4\pi}}P_\ell(\cos\theta).
\end{equation}
Because the coordinates $(r,\theta)$ and $(\tilde r,\tilde \theta)$ differ by a perturbation,
\begin{equation}
r = \tilde r+\delta r,\quad \theta=\tilde \theta+\delta\theta.
\end{equation}
we may take $H_{0,\ell}$ and $\mathcal K_\ell$ to be functions of $r$ rather than $\tilde r$, and $Y_\ell$ to be a function of $\theta$. 

To build the dictionary between the Weyl and Regge--Wheeler formalisms, we compute spacetime scalars in each one and match. The simplest scalars available to us are the $g_{tt}$ and $g_{\phi\phi}$ metric components, which are the norms of the two Killing vectors. Comparing these between \cref{eq:RW-metric,eq:met-GH-lin} we see that
\begin{subequations}\label{eq:RWmatch}
\begin{align}
f(r)(1-\hat\psi) &=f(\tilde r)\left(1-\displaystyle\sum_\ell H_{0,\ell}(r)Y_{\ell}(\theta)\right),\label{eq:RWmatch1}\\
\Delta(r)\sin^2\theta &= \Delta(\tilde r) \sin^2\tilde\theta\left(1+\displaystyle\sum_\ell \kappa_{\ell}(r)Y_{\ell}(\theta)\right),\label{eq:RWmatch2}
\end{align}
\end{subequations}
where for convenience we have performed the field redefinition
\begin{equation}
\kappa \equiv \mathcal K - H_0.
\end{equation}
Noting that
\begin{equation}
f(\tilde r) = f(r)\left(1-\frac\rs\Delta\delta r\right),
\end{equation}
we may linearize \cref{eq:RWmatch1} to relate the Regge--Wheeler metric perturbation $H_0$ to the linearized Geroch--Hartle distorted potential,
\begin{equation}
\hat\psi = \displaystyle\sum H_{0,\ell}(r)Y_\ell(\theta) + \frac\rs\Delta\delta r.
\label{eq:FieldsMatch}
\end{equation}
From \cref{eq:RWmatch2} we find $\rho$ in Regge--Wheeler coordinates,
\begin{equation}
\rho = \sqrt{\Delta(\tilde{r})}\sin\tilde\theta\left(1+\frac12\displaystyle\sum_\ell \kappa_{\ell}(r)Y_{\ell}(\theta)\right).
\end{equation}
We can then compute $z(\tilde r,\tilde\theta)$ by linearizing and integrating the defining relation $\dd z = -\star\dd\rho$. Note that this is not entirely trivial as we must compute the perturbation of the 2D Hodge star. The result is
\begin{equation}
z = \left( \tilde r-\frac\rs 2 \right)\cos\tilde{\theta} + \sum_\ell \frac{\sin\theta}{2\ell(\ell+1)}\frac{\dd Y_\ell}{\dd\theta}\frac{\dd (\Delta\kappa_\ell)}{\dd r}. \label{eq:zRW}
\end{equation}
The inverse $\ell(\ell+1)$ factor reflects the nonlocal relation between $\rho$ and $z$.
Finally we linearize \cref{eq:sch-weyl-coord} to obtain
\begin{subequations}
\begin{align}
\delta r &= \sum_{\ell}\frac12 \mathrm{e}^{2\gamma_\sch}\sin^2\theta\left(\frac12\Delta'\kappa_\ell Y_\ell + \frac{\cot\theta}{\ell(\ell+1)}\frac{\dd(\Delta\kappa_\ell)}{\dd r}\frac{\dd Y_\ell}{\dd\theta}\right), \\
\delta \cos\theta &= \sum_{\ell}\frac12 \mathrm{e}^{2\gamma_\sch}\sin^2\theta\left(\frac{\Delta'}{2\Delta}\frac{\sin\theta }{\ell(\ell+1)}\frac{\dd(\Delta\kappa_\ell)}{\dd r}\frac{\dd Y_\ell}{\dd\theta}-\cos\theta \kappa_\ell Y_\ell \right).
\end{align} 
\end{subequations}
Recall that $\Delta'\equiv\dd\Delta/\dd r=2r-\rs$. As a consistency check we note these satisfy \cref{eq:RWmatch2}. 

It is instructive to study the linearized Einstein equations and their
solutions in both the Weyl and BHPT/Regge--Wheeler frameworks.
By linearizing the equation $\Box_2\rho=0$ in Regge--Wheeler gauge we find a second-order equation for $\kappa_\ell$,
\begin{equation}
\frac{\dd^2(\Delta \kappa_\ell)}{\dd r^2}=\ell(\ell+1)\kappa_\ell.\label{eq:linRadEom}
\end{equation}
By linearizing the 2D Einstein equation, or equivalently \cref{eq:lambda-main}, we have a constraint between $H_0$ and $\kappa$,
\begin{equation}
\frac{\dd \kappa_\ell}{\dd r}=\frac{\rs}{\Delta} H_{0,\ell},
\end{equation}
and by linearizing the $\psi$ equation \eqref{eq:psi-eom} we obtain an equation of motion for $H_0$ sourced by $\kappa$,
\begin{equation}
 \frac{\dd}{\dd r}\left(\Delta \frac{\dd H_{0,\ell}}{\dd r} \right) - \ell(\ell+1) H_{0,\ell} = \rs \frac{\dd \kappa_\ell}{\dd r}, 
\end{equation}
By combining the first two equations we may integrate out $\kappa$ to obtain an equation for $H_0$ alone \cite{Hinderer:2007mb,Riva:2023rcm},
\begin{equation}
 \frac{\dd}{\dd r}\left(\Delta \frac{\dd H_{0,\ell}}{\dd r} \right) - \frac{\rs^2+\ell(\ell+1)\Delta}\Delta H_{0,\ell} = 0.
\end{equation}
The horizon-regular solution is
\begin{align}
H_{0,\ell} = \frac{1}{\ell!} \mathcal{E}_\ell P_\ell^2(x), \qquad \kappa_\ell = \frac{1}{\ell!} \mathcal{E}_\ell\frac\rs{\sqrt\Delta}P_\ell^1(x),
\end{align}
where $\mathcal E_\ell$ is a constant and the functional dependence is, as usual,
\begin{equation}
x\equiv \frac{\Delta'}\rs = \frac{2r}\rs-1.
\end{equation}

Notice that the solutions for $H_{0,\ell}$ are $m=2$ associated Legendre polynomials in $\Delta'/\rs$, while, as we have seen, modes of $\hat\psi$ are $m=0$ Legendre polynomials, cf. \cref{eq:hatpsiphysicalsolution}. This seemingly-contradictory behavior is in fact explained entirely by \cref{eq:FieldsMatch} and the observation that the multipole expansions in the Weyl and Regge--Wheeler coordinates are not quite the same. For instance, plugging the $\ell=2$ Regge--Wheeler solution into \cref{eq:FieldsMatch} we find that the corresponding $\hat\psi$ is a sum of monopole and quadrupole contributions,
\begin{align}
\hat\psi &= H_{0,2}(r)Y_2(\theta) + \frac\rs\Delta\delta r  \nonumber\\
&= \sqrt{\frac5{4\pi}}\frac3{2\rs^2}\left[2\Delta-\rs^2+\left(6\Delta+\rs^2\right)\cos2\theta\right] \nonumber\\
&= 2\sqrt{\frac5{4\pi}}\left(P_2(x)P_2(\cos\theta)-P_0(x)P_0(\cos\theta)\right),
\end{align}
that is, an $\ell=2$ Regge--Wheeler solution corresponds to a
distorted black hole with $a_2 = -a_0 = 2\sqrt{5/(4\pi)}$. In general
a Regge--Wheeler mode with $\ell$ even (odd) will correspond to a particular sum of distorted black holes at levels $n$ where $n\leq \ell$ is also even (odd).
We present the dictionary up through $\ell=6$
(on the left is $\ell$ in Regge--Wheeler, and on the right are
the corresponding $a_i$ coefficients as in
\cref{eq:hatpsiphysicalsolution}, labeled as $a_0, a_1,  \dots$):
\begin{subequations}
\begin{align}
\ell=2:&& a_i &= 2\sqrt{\frac5{4\pi}}(-1,0,1),\\
\ell=3:&& a_i &= 6\sqrt{\frac7{4\pi}}(0,-1,0,1),\\
\ell=4:&& a_i &= 2\sqrt{\frac9{4\pi}}(-1,0,-5,0,6),\\
\ell=5:&& a_i &= 2\sqrt{\frac{11}{4\pi}}(0,-3,0,-7,0,10),\\
\ell=6:&& a_i &= 2\sqrt{\frac{13}{4\pi}}(-1,0,-5,0,-9,0,15).
\end{align}
\end{subequations}

\section{Axisymmetric operators}
\label{sec:axiop}

{In this appendix, we elaborate on \cref{Sintalt} and show how one can systematically construct $\tilde S_\mathrm{int}$.
  It is useful to start with a concrete example. Consider the $\ell=2$ tidal field
  $E_{i_1 i_2}$. This is a $3 \times 3$ symmetric traceless
  matrix. Its five independent components can be expressed in many
  different ways. One way is to use the five matrices $c_{i_1 i_2}^m$
  (for $m=-2, -1, 0, 1, 2$) associated with
  the $\ell=2$ spherical harmonics, defined by
  \begin{equation}
  r^2 Y_{2 m} (\theta, \phi) \equiv \sum_{i_1, i_2} c_{i_1 i_2}^m 
  x^{i_1} x^{i_2} \, ,
\end{equation}
where $r^2 = \sum_{i} x^i {}^2$. One can write
\begin{equation}
  E_{i_1 i_2} = \sum_m {\mathcal E}^m c_{i_1 i_2}^m \, .
\end{equation}
It is worth noting that this decomposition into ${\mathcal E}^m$'s is
general, and does not assume linear theory or the equation of motion.
The coefficients ${\mathcal E}$'s can in general be functions of space. {We will assume below that their space dependence, if present, respects axisymmetry, which will be the case for the problem under consideration.} We will be
particularly interested in $m=0$, due to its association with
axisymmetry. (We will see more explicity the role $m=0$ plays
when we perform the matching exercise between the EFT
and the exact Weyl solution.) It is thus useful to define a projection:
\begin{equation}
  E^{\cal P}_{i_1 i_2} \equiv
{{\cal P}_{\langle i_1}}^{\! i_1'} {{\cal P}_{i_2 \rangle}}^{\! i_2' } E_{i_1' i_2' } \equiv
  {{\cal P}_{i_1}}^{\! i_1'} {{\cal P}_{i_2}}^{\! i_2'} E_{i_1' i_2'} - {1\over
  3} \delta_{i_1 i_2}  {\cal P}^{i i_1'} {{\cal P}_{i}}^{ i_2'}
E_{i_1' i_2'} \, ,
\end{equation}
where ${\cal P}_{i_1 i_2} \equiv \delta_{i_1 3} \delta_{i_2 3}$.
It can be verified that $E^{\cal P}_{i_1 i_2} = {\mathcal E}^0 c^0_{i_1
  i_2}$. The projection can be extended to any tidal field with
$\ell$ number of indices: $E^{\cal P}_{i_1 \cdots i_\ell} \equiv
{{\cal P}_{\langle i_1}}^{\! i_1'} \cdots {{\cal P}_{i_\ell \rangle}}^{\! i_\ell'} E_{i_1'
  \cdots i_\ell' }$.}
{By construction, any contraction of an arbitrary number of $E^{\cal P}$'s, with $\ell$ remaining free indices, is proportional to $E^{\cal P}_{i_1 \cdots i_\ell}$ itself, after all traces are subtracted.}

{With the projection defined, we can look more closely at
operators in \eqref{Sintexp} where multiple contractions are
possible. For instance, consider the operator with 6 $E$'s, each with
$\ell=2$. There are two possible contractions:
$({\rm Tr} E^3)^2 = (\sum_{ijk} E_{ij} E_{jk} E_{ki})^2$ and
$({\rm Tr} E^2)^3 = (\sum_{ij} E_{ij} E_{ji})^3$. (It can be shown
other contractions can be rewritten in terms of them~\cite{Bern:2020uwk}.)
Once we perform projections though, these two possible contractions
are in fact proportional to each other, that is to say,
$({\rm Tr} E^{\cal P}  {}^2)^3 = 6 ({\rm Tr} E^{\cal P}  {}^3)^2$. For this reason,
at the level of $6$ $E$'s and $\ell=\ell_1 = \ell_2 = \ell_3 = \ell_4
= \ell_5 = 2$, a convenient way to express the Love number operators
would be:
\begin{equation}
S_{\rm int} \supset \int {\rm d}\tau \left( \lambda^{(5)}_{222222} ({\rm Tr}
E^{\cal P}
{}^2)^3 + \tilde\lambda^{(5)}_{222222} \left[ ({\rm Tr} E^2)^3
  - 6 ({\rm Tr} E^3)^2 \right]  \right)\, .
\label{Sint6th}
\end{equation}
{This is convenient because, when evaluated on an axisymmetric  configuration in the equations of motion, the second operator vanishes identically. Indeed, the variation of the second term in \eqref{Sint6th} yields $\delta S_{\rm int}= 6\tilde\lambda^{(5)}_{222222} [  ({\rm Tr} E^2)^2 E_{ij}
  - 6 ({\rm Tr} E^3) {E_{i}}^kE_{kj} ] \delta E^{ij}$, which is zero when the $E$'s in parenthesis are of the form ${\cal E}^0c^0_{ij}$, by virtue of the tracelessness of $\delta E^{ij}$.
 The
coefficient $\tilde\lambda^{(5)}_{222222}$ is 
therefore unconstrained by matching with the axisymmetric Weyl
solution.}
The coefficient of the first operator in \eqref{Sint6th} is what we will be able to deduce by
matching. Note that we could also dispense with the projection in the
first term, and write the action simply as:
\begin{equation}
S_{\rm int} \supset \int {\rm d}\tau \left( \lambda^{(5)}_{222222} ({\rm Tr} E^2)^3 + \tilde\lambda^{(5)}_{222222} \left[ ({\rm Tr} E^2)^3
  - 6 ({\rm Tr} E^3)^2 \right] \right) \, .
\end{equation}
After all, the basis in which we express the Love number operators is
up to us. The important point is that the second operator {gives a vanishing contribution in the equations of motion} by
definition for axisymmetric configurations, while the first does not.}

{The above example illustrates a general principle: for each
  distinct multiplet $(\ell \ell_1 \cdots \ell_n)$, associated with $n+1$
  number of fields, there will be one Love number operator whose coupling we
  will be able to deduce by matching with the Weyl solution.}
  {All the other independent operators at the same perturbative order in the EFT can be arranged  in a way that they do not contribute to the matching with axisymmetric field configurations. 
All in all, the action can be expressed in general as  in  \cref{Sintalt}.}

\section{Horizontal ladder symmetries and the Wronskian}
\label{sec:ladders}

In this section, we discuss some properties of the symmetries presented in section~\ref{sec:sym}. We will first briefly review some properties of the ladder symmetries \eqref{Ddef}. 
We will then explain how the horizontal symmetries are connected to the conserved Wronskian and the Geroch group.

Let us start again from the $\hat \psi$'s equation \eqref{psieomSch3}, which we can rewrite for convenience more compactly as
\begin{equation}
E_\ell\hat{\psi}_\ell = 0 ,
\qquad\quad
E_\ell \equiv \partial_r\left(\Delta\partial_r\right)-\ell(\ell+1).
\label{eq:Ell}
\end{equation}
Equivalently, one can work at the level of the Lagrangian  from which \cref{eq:Ell} is derived, i.e.,
\begin{align}
\label{eq:Lellapp}
\mathcal L_\ell &= -\frac12\left(\Delta\hat\psi_\ell'^2+\ell(\ell+1)\hat\psi_\ell^2\right) \nonumber \\
&= \frac12\hat\psi_\ell E_\ell\hat\psi_\ell ,
\end{align}
for each mode $\hat{\psi}_\ell$.

It is instructive to define the ``Hamiltonian''
\begin{equation}
H_\ell \equiv -\Delta E_\ell = - \Delta \partial_r\left(\Delta\partial_r\right) +\ell(\ell+1)\Delta,
\end{equation}
such that a physical mode $\hat\psi_\ell $ which solves the equation \eqref{eq:Ell} satisfies $H_\ell\hat\psi_\ell=0$.
It is straightforward to check that the ``vertical ladder operators" \eqref{Ddef} \cite{Hui:2021vcv,Compton:2020cjx,Hui:2022vbh},
\begin{equation}
D_\ell^+ = -\Delta \partial_r -\frac{\ell+1}{2}\Delta',
\qquad
D_\ell^- = \Delta\partial_r -\frac\ell2\Delta',
\end{equation}
satisfy the operator identities
\begin{subequations}
\label{eq:HDDH}
\begin{align}
H_{\ell+1}D_\ell^+ &= D_\ell^+H_\ell,\\
H_{\ell-1}D_\ell^- &= D_\ell^-H_\ell.
\end{align}
\end{subequations}
These relations make it clear that $D_\ell^+$ and $D_\ell^-$ represent  raising and lowering operators in the harmonic number $\ell$: $D_\ell^+$ and $D_\ell^-$ acting on a solution for a given $\ell$ yield solutions at $\ell+1$ and $\ell-1$ levels, respectively. 

The vertical ladder operators satisfy some useful intertwining relations \cite{Hui:2021vcv}. In particular, they can be used to rewrite the Hamiltonian operator $H_\ell$ as
\begin{align}
H_\ell&=D_{\ell-1}^+D_\ell^- - \frac{\ell^2}4\rs^2.\label{eq:H-inter}
\end{align}

In addition to the vertical ladders, we can  construct ``horizontal" symmetry operators,  which act on a single $\ell$ multipole.
They can be defined recursively in terms of the $D^\pm_\ell$ operators as \cite{Hui:2021vcv}
\begin{align}\label{eq:higher-ell-Q}
Q_0 & = D_0^-,\\
Q_\ell &= D_{\ell-1}^+Q_{\ell-1} D^-_\ell \nonumber \\
&=D_{\ell-1}^+\cdots D_0^+D_0^-\cdots D_\ell^-,
\end{align}
such that they satisfy the commutation relations $[Q_\ell,H_\ell]=0$. 
It is straightforward to check that $Q_\ell$ leave the Lagrangians \eqref{eq:Lellapp} invariant, up to total derivative terms.

It can be shown that their associated conserved charges (conserved in
the sense of being $r$-independent) take the form of the Wronskian
squared \cite{Compton:2020cjx,BenAchour:2022uqo}.
For instance, at $\ell=0$, the Noether charge corresponding
to the symmetry $\delta\hat \psi_0 = Q_0 \hat \psi_0$ turns out to be
$(\Delta \partial_r \hat \psi_0)^2$ (the conservation of which can
be easily checked), and $\Delta \partial_r \hat \psi_0$
can be thought of as the Wronskian between the regular solution
$\hat \psi_0 = 1$ and a generic $\hat \psi_0$, i.e.~$W(1, \hat\psi_0) \equiv 1
\Delta \partial_r \hat\psi_0 - \hat \psi_0 \Delta \partial_r 1 = \Delta
\partial_r \hat\psi_0$. Another way of saying this, in the language of
charge generating symmetry, is
$\delta \hat\psi_\ell \equiv \{W^2 , \hat\psi_\ell \} =
-2 \hat\psi_\ell^{\rm reg} W$, where  $W$ represents
$W[\hat\psi_\ell^{\rm reg} , \hat\psi_\ell]$,
$\hat\psi_\ell^{\rm reg}$ denotes the regular solution while
$\hat\psi$ denotes a generic field configuration, and $\{\,\}$ represents
the Poisson bracket.\footnote{Here,
$\{ A, B \} \equiv \frac{\delta A}{\delta \hat\psi_\ell} \frac{\delta B}{\delta \Delta\partial_r \hat\psi_\ell} - \frac{\delta A}{\delta \Delta
  \partial_r \hat\psi_\ell} \frac{\delta B}{\delta \hat\psi_\ell}$.
}
The symmetry transformation $\delta\hat\psi_\ell =
\psi_\ell^{\rm reg} W$
(dropping the factor of $-2$) looks rather different from
$\delta\hat\psi_\ell = Q_\ell \hat\psi_\ell$, but it can be
shown they are equivalent. Below is a little detour to
establish this fact. Readers not interested in the proof can skip to
below \cref{equivalencedS}. 

It is convenient to define
\begin{equation}
Q'_\ell \hat\psi_\ell \equiv \hat \psi_\ell^{\rm reg}
W[\hat\psi_\ell^{\rm reg} , \hat\psi_\ell] \, ,
\end{equation}
where $\hat\psi_\ell^{\rm reg} \equiv D^+_{\ell-1} ... D^+_0 1 =
(-\rs/2)^\ell \ell ! P_\ell(\Delta'/\rs)$. That $Q_0 = Q_0'$ is simple
to see. Let us look at $\ell=1$:
\begin{align}
Q_1 &= D_0^+D_0^-D_1^- \nonumber\\
&= [D_0^+,  D_r ]D_1^-+D_rD_0^+D_1^- \nonumber\\
&= \Delta D_1^-+D_r\left(H_1+\frac{\rs^2}4\right) \nonumber\\
&= D_rH_1 + Q_1' \, ,
\end{align}
where we have defined $D_r \equiv \Delta \partial_r$. Thus, on shell,
$Q_1 = Q_1'$. But one can make a stronger statement.
Both $Q_1\hat\psi_1$ and $Q_1' \hat\psi_1$ are good
off-shell symmetries of the action, and their induced variations
of the action differ only by a total derivative:
\begin{align}
\Delta\delta S_1 \equiv \delta S_1-\delta' S_1 &= \int \dd r (E_1\hat\psi_1)(D_rH_1\hat\psi_1) \nonumber\\
&= -\int \dd r (H_1\hat \psi_1)(\partial_rH_1\hat\psi_1) \nonumber\\
&=-\frac12\int\dd r\partial_r(H_1\hat \psi_1)^2.
\end{align}
Thus we may regard $Q_1$ and the reduced-order $Q_1'$ as equivalent,
$Q_1\approx Q_1'$.
Now we may proceed recursively. Let us assume that, at order $\ell$,
we have the same sort of equivalence we found for $\ell=0,1$, i.e.~$Q_\ell \approx Q'_\ell$.
Then we can construct the order $\ell+1$ operator with three
derivatives rather than $2\ell+3$ of them, and proceed analogously to
the $\ell=1$ case, 
\begin{align}
Q_{\ell+1} &\approx D_\ell^+Q'_\ell D_{\ell+1}^- \nonumber\\
&= [D_\ell^+,Q'_\ell] D_{\ell+1}^- + Q'_\ell D_\ell^+ D_{\ell+1}^- \nonumber\\
&=Q'_\ell H_{\ell+1} + [D_\ell^+,Q'_\ell] D_{\ell+1}^- +  \frac{(\ell+1)^2}4\rs^2Q'_\ell\nonumber\\
&= \hat\psi_\ell^{\rm reg} D_r(\hat\psi_\ell^{\rm reg} H_{\ell+1}) + Q'_{\ell+1}.
\end{align}
The difference in variations of the action is a total derivative,
\begin{align}
\label{equivalencedS}
\Delta\delta S_{\ell+1} &= \int\dd r
                          (E_{\ell+1}\hat\psi_{\ell+1})(Q_{\ell+1}-Q'_{\ell+1})\hat\psi_{\ell+1} \nonumber \\
&= - \int \dd r \hat\psi_\ell^{\rm reg} (H_{\ell+1}\hat \psi_{\ell+1})
                                                                                                                \partial_r(\hat\psi_\ell^{\rm
                                                                                                                reg}
                                                                                                                H_{\ell+1}\hat\psi_{\ell+1})
                                                                                                                \nonumber \\
&= -\frac12\int \dd r \partial_r\left(H_{\ell+1}\hat\psi_{\ell+1}\right)^2,
\end{align}
establishing equivalence between $Q_{\ell+1}$ and $Q'_{\ell+1}$. Having already shown this equivalence for $\ell=0,1$, it follows that it holds for all $\ell$.

Having established that the horizontal symmetries and symmetries
generated by the Wronskian squared are equivalent, it is natural to ask:
what about symmetries generated by the Wronskian alone? 
While the Wronskian squared generates a linear transformation, in the
sense that $\delta \hat\psi_\ell = \{W^2 , \hat\psi_\ell\}$ is proportional
to $\hat\psi_\ell$,  the Wronksian by itself generates a nonlinear
transformation,
in the sense that $\delta \hat\psi_\ell = \{W , \hat\psi_\ell\}$ does not
depend on $\hat\psi_\ell$. Not surprisingly, the latter is simply shifting
$\hat\psi$ by a solution. It is shown by \cite{BenAchour:2022uqo}
that the full set of symmetry transformations generated by $W$ and
$W^2$, using both the regular and the irregular solutions (i.e., two
different Wronksians and their products), form
an algebra which is the semi-direct product of
$\mathfrak{sl}(2,\mathbb R)$
and a Heisenberg algebra, with a central charge given by the Wronskian
between the regular and irregular solutions.
A natural infinite-dimensional extension would be to include also
$W^3$, $W^4$ and so on.

\section{Geroch symmetry: a primer}
\label{app:geroch}

The fact that the Geroch group is infinite-dimensional is another way of saying that general relativity for static and axisymmetric spacetimes is integrable. As for other integrable systems, the Laplace equation \eqref{eq:psi-eom} is equivalent to a system of first-order equations known as a \emph{Lax pair},\footnote{In forms notation, $\dd X = (1+N_+)\dd \psi + N_-\star\dd\psi$.}
\begin{subequations}\label{eq:lax}
\begin{align}
\label{eq:lax-rho}\partial_\rho X &= (1+N_+)\partial_\rho\psi + N_-\partial_z\psi,\\
\label{eq:lax-z}\partial_z X &= (1+N_+)\partial_z\psi - N_-\partial_\rho\psi.
\end{align}
\end{subequations}
Here $X(\rho,z;w)$ is a field which depends on the spatial coordinates as well as a constant \emph{spectral parameter} $w$, while $N_\pm(\rho,z;w)$ are specified functions of space and of $w$,
\begin{subequations}
\begin{align}
N_+ &= \frac{z-w}{\sqrt{\rho^2+(w-z)^2}},\label{eq:Np}\\
N_- &= \frac{\rho}{\sqrt{\rho^2+(w-z)^2}}.\label{eq:Nm}
\end{align}
\end{subequations}
By taking a $z$ derivative of \cref{eq:lax-rho} and a $\rho$ derivative of \cref{eq:lax-z} we obtain a compatibility condition which is precisely \cref{eq:psi-eom}.

The functions $N_\pm(\rho,z;w)$ are rather interesting. It is straightforward to check that they obey the duality relation\footnote{In forms, $\dd N_+ = \rho\star\dd(N_-/\rho).$}
\begin{equation}
\label{eq:dual}
\partial_\rho N_+ = \partial_z N_-, \qquad \partial_z N_+ = -\rho\partial_\rho\left(\frac{N_-}\rho\right),
\end{equation}
and that their squares sum to unity,
\begin{equation}
N_+^2+N_-^2 = 1.
\end{equation}
It follows from \cref{eq:dual} that the function
\begin{equation}
\Psi \equiv \frac{N_-}{\rho} = \frac 1{\sqrt{\rho^2+(w-z)^2}}\label{eq:Psi}
\end{equation}
is itself a solution to \cref{eq:psi-eom} for any value of the spectral parameter $w$.\footnote{This is most easily seen in the language of differential forms, where it is just a consequence of $\dd^2=0$, recalling that the Laplace equation \eqref{eq:psi-eom} may be written as $\dd(\rho\star\dd\psi)=0$.} In fact this solution is a \emph{generating function} for growing modes, as can be seen by expanding around $w=\infty$,
\begin{subequations}\label{eq:Npn}
\begin{align}
N_+(\rho,z;w) &= -1+ \displaystyle\sum_{n=1}^\infty N_+^{(n)}(\rho,z) w^{-n-1},\\
N_-(\rho,z;w) &= \displaystyle\sum_{n=0}^\infty N_-^{(n)}(\rho,z) w^{-n-1},
\end{align}
\end{subequations}
where, in terms of the polar coordinates $(\rho,z)=(\mathcal R\sin\vartheta,\mathcal R\cos\vartheta)$, the coefficients $N_\pm^{(n)}$ are\footnote{We use the notation $N_\pm$ following \cite{Gurlebeck:2012mb,Gurlebeck:2013eia,Gurlebeck:2015xpa}, where coefficients $N_\pm^{(n)}$ were defined that are equivalent to our expressions up to an overall factor of $-2$.}
\begin{subequations}
\begin{align}
N_+^{(n)}&= -\frac{\rho}{n+1}\mathcal{R}^n P_n^1(\cos\vartheta),\\
N_-^{(n)} &=\rho \mathcal{R}^n P_n(\cos\vartheta).
\end{align}
\end{subequations}
We may now make precise our claim that $\Psi$ is a generating function for growing solutions,
\begin{align}\label{eq:Psi-sep}
\Psi = \frac 1{\sqrt{\rho^2+(w-z)^2}}= \displaystyle\sum_{n=0}^\infty \frac{\mathcal R^n P_n (\cos\vartheta)}{ w^{n+1}}.
\end{align}
Note that the theorem of \cite{Gurlebeck:2015xpa}, which is close in spirit to our result, follows from integrating \cref{eq:lax}, and relies deeply on the properties of the $N_\pm$.

Expositions of the Geroch group typically rely on the presence of the field $\chi$ which we have set to zero, cf.~\cref{app:dimred}. The usual story is that when we reduce to $D=2$, in addition to the Ehlers $\mathrm{SL}(2,\mathbb R)/\mathrm{SO}(2)$ group acting on $(\psi,\chi)$, we have an $\mathrm{SL}(2,\mathbb R)/\mathrm{SO}(1,1)$ Matzner--Misner symmetry which results from not dualizing the 3-vector $A_i$ into the scalar $\chi$. In the Matzer--Misner description we instead act on the $\phi$ component of $A_i$, which is related to $\chi$ by
\begin{equation}
\star\dd A_\phi = \rho e^{2\phi}\dd\chi.
\end{equation}
Because of this nonlocal relation, acting an element of the Matzner--Misner group on $\chi$, or acting an Ehlers element on $A_\phi$, requires the introduction of a new, nonlocally related potential. This process continues \emph{ad infinitum}, and an infinite tower of fields is required to locally describe the action of these groups. The infinite-dimensional Geroch group is the result of this failure of the Ehlers and Matzner--Misner groups to commute. In the static case considered in this work, we lack most of this structure, but there is a remnant, as $\psi$ becomes shift-symmetric and has a dual scalar $\bar\psi$ defined by
\begin{equation}
\dd\bar\psi = \rho\star\dd\psi.
\end{equation}

In the general picture, the fields $(\psi,\chi)$ are gathered into an $\mathrm{SL}(2,\mathbb R)$ matrix-valued coset representative $\mathcal V$, and the field $X$ appearing in the Lax equation \eqref{eq:lax} is also an $\mathrm{SL}(2,\mathbb R)$ matrix. The infinitesimal action of the Geroch symmetry is \cite{Lu:2007jc}
\begin{equation}
\delta\mathcal V = \Psi \mathcal V \eta + \delta h \mathcal V,
\end{equation}
where $\delta h$ is a compensating transformation to restore the original $\mathfrak{so}(2)$ gauge choice, and $\eta$ is a $w$-dependent infinitesimal factor, obtained by conjugating an $\mathfrak{sl}(2,\mathbb R)$-valued constant infinitesimal parameter $\epsilon$ by $X$,
\begin{equation}
\eta = X\epsilon X^{-1}.
\end{equation}
For the static case, everything is a singlet under $\mathrm{SL}(2,\mathbb R)$ and we have functions rather than matrices. This implies $\eta=\epsilon$, and following \cite{Lu:2007jc} we choose $\epsilon=w$ so that $\delta\psi\to1$ at $w\to\infty$. Thus we find $\delta\psi = w \Psi$, cf. \cref{eq:perry-sym-static}.

\addcontentsline{toc}{section}{References}
\bibliographystyle{utphys}
{ %
\bibliography{biblio}

\providecommand{\href}[2]{#2}\begingroup\raggedright\begin{thebibliography}{10}

\bibitem{Chandrasekhar:1985kt}
S.~Chandrasekhar,
  \href{http://dx.doi.org/10.1093/oso/9780198503705.001.0001}{{\em {The
  mathematical theory of black holes}}}.
\newblock 1985.

\bibitem{Israel:1967wq}
W.~Israel, ``{Event horizons in static vacuum space-times},''
  \href{http://dx.doi.org/10.1103/PhysRev.164.1776}{{\em Phys. Rev.} {\bfseries
  164} (1967) 1776--1779}.

\bibitem{Carter:1968rr}
B.~Carter, ``{Global structure of the Kerr family of gravitational fields},''
  \href{http://dx.doi.org/10.1103/PhysRev.174.1559}{{\em Phys. Rev.} {\bfseries
  174} (1968) 1559--1571}.

\bibitem{Carter:1971zc}
B.~Carter, ``{Axisymmetric Black Hole Has Only Two Degrees of Freedom},''
  \href{http://dx.doi.org/10.1103/PhysRevLett.26.331}{{\em Phys. Rev. Lett.}
  {\bfseries 26} (1971) 331--333}.

\bibitem{Wald:1971iw}
R.~M. Wald, ``{Final states of gravitational collapse},''
  \href{http://dx.doi.org/10.1103/PhysRevLett.26.1653}{{\em Phys. Rev. Lett.}
  {\bfseries 26} (1971) 1653--1655}.

\bibitem{Hartle:1971qq}
J.~B. Hartle, ``{Long-range neutrino forces exerted by kerr black holes},''
  \href{http://dx.doi.org/10.1103/PhysRevD.3.2938}{{\em Phys. Rev. D}
  {\bfseries 3} (1971) 2938--2940}.

\bibitem{Bekenstein:1971hc}
J.~D. Bekenstein, ``{Nonexistence of baryon number for static black holes},''
  \href{http://dx.doi.org/10.1103/PhysRevD.5.1239}{{\em Phys. Rev. D}
  {\bfseries 5} (1972) 1239--1246}.

\bibitem{Fackerell:1972hg}
E.~D. Fackerell and J.~R. Ipser, ``{Weak electromagnetic fields around a
  rotating black hole},'' \href{http://dx.doi.org/10.1103/PhysRevD.5.2455}{{\em
  Phys. Rev. D} {\bfseries 5} (1972) 2455--2458}.

\bibitem{Price:1972pw}
R.~H. Price, ``{Nonspherical Perturbations of Relativistic Gravitational
  Collapse. II. Integer-Spin, Zero-Rest-Mass Fields},''
  \href{http://dx.doi.org/10.1103/PhysRevD.5.2439}{{\em Phys. Rev. D}
  {\bfseries 5} (1972) 2439--2454}.

\bibitem{Bekenstein:1995un}
J.~Bekenstein, ``{Novel ``no-scalar-hair" theorem for black holes},''
  \href{http://dx.doi.org/10.1103/PhysRevD.51.R6608}{{\em Phys. Rev. D}
  {\bfseries 51} no.~12, (1995) 6608}.

\bibitem{Hui:2012qt}
L.~Hui and A.~Nicolis, ``{No-Hair Theorem for the Galileon},''
  \href{http://dx.doi.org/10.1103/PhysRevLett.110.241104}{{\em Phys. Rev.
  Lett.} {\bfseries 110} (2013) 241104},
  \href{http://arxiv.org/abs/1202.1296}{{\ttfamily arXiv:1202.1296 [hep-th]}}.

\bibitem{Fang:2005qq}
H.~Fang and G.~Lovelace, ``{Tidal coupling of a Schwarzschild black hole and
  circularly orbiting moon},''
  \href{http://dx.doi.org/10.1103/PhysRevD.72.124016}{{\em Phys. Rev. D}
  {\bfseries 72} (2005) 124016},
  \href{http://arxiv.org/abs/gr-qc/0505156}{{\ttfamily arXiv:gr-qc/0505156}}.

\bibitem{Damour:2009vw}
T.~Damour and A.~Nagar, ``{Relativistic tidal properties of neutron stars},''
  \href{http://dx.doi.org/10.1103/PhysRevD.80.084035}{{\em Phys. Rev. D}
  {\bfseries 80} (2009) 084035},
  \href{http://arxiv.org/abs/0906.0096}{{\ttfamily arXiv:0906.0096 [gr-qc]}}.

\bibitem{Binnington:2009bb}
T.~Binnington and E.~Poisson, ``{Relativistic theory of tidal Love numbers},''
  \href{http://dx.doi.org/10.1103/PhysRevD.80.084018}{{\em Phys. Rev. D}
  {\bfseries 80} (2009) 084018},
  \href{http://arxiv.org/abs/0906.1366}{{\ttfamily arXiv:0906.1366 [gr-qc]}}.

\bibitem{Goldberger:2004jt}
W.~D. Goldberger and I.~Z. Rothstein, ``{An Effective field theory of gravity
  for extended objects},''
  \href{http://dx.doi.org/10.1103/PhysRevD.73.104029}{{\em Phys. Rev. D}
  {\bfseries 73} (2006) 104029},
  \href{http://arxiv.org/abs/hep-th/0409156}{{\ttfamily arXiv:hep-th/0409156}}.

\bibitem{Goldberger:2005cd}
W.~D. Goldberger and I.~Z. Rothstein, ``{Dissipative effects in the worldline
  approach to black hole dynamics},''
  \href{http://dx.doi.org/10.1103/PhysRevD.73.104030}{{\em Phys. Rev. D}
  {\bfseries 73} (2006) 104030},
  \href{http://arxiv.org/abs/hep-th/0511133}{{\ttfamily arXiv:hep-th/0511133}}.

\bibitem{Kol:2011vg}
B.~Kol and M.~Smolkin, ``{Black hole stereotyping: Induced gravito-static
  polarization},'' \href{http://dx.doi.org/10.1007/JHEP02(2012)010}{{\em JHEP}
  {\bfseries 02} (2012) 010}, \href{http://arxiv.org/abs/1110.3764}{{\ttfamily
  arXiv:1110.3764 [hep-th]}}.

\bibitem{Gurlebeck:2015xpa}
N.~G\"urlebeck, ``{No-hair theorem for Black Holes in Astrophysical
  Environments},'' \href{http://dx.doi.org/10.1103/PhysRevLett.114.151102}{{\em
  Phys. Rev. Lett.} {\bfseries 114} no.~15, (2015) 151102},
  \href{http://arxiv.org/abs/1503.03240}{{\ttfamily arXiv:1503.03240 [gr-qc]}}.

\bibitem{Hui:2020xxx}
L.~Hui, A.~Joyce, R.~Penco, L.~Santoni, and A.~R. Solomon, ``{Static response
  and Love numbers of Schwarzschild black holes},''
  \href{http://dx.doi.org/10.1088/1475-7516/2021/04/052}{{\em JCAP} {\bfseries
  04} (2021) 052}, \href{http://arxiv.org/abs/2010.00593}{{\ttfamily
  arXiv:2010.00593 [hep-th]}}.

\bibitem{Hui:2021vcv}
L.~Hui, A.~Joyce, R.~Penco, L.~Santoni, and A.~R. Solomon, ``{Ladder symmetries
  of black holes. Implications for love numbers and no-hair theorems},''
  \href{http://dx.doi.org/10.1088/1475-7516/2022/01/032}{{\em JCAP} {\bfseries
  01} no.~01, (2022) 032}, \href{http://arxiv.org/abs/2105.01069}{{\ttfamily
  arXiv:2105.01069 [hep-th]}}.

\bibitem{Hui:2022vbh}
L.~Hui, A.~Joyce, R.~Penco, L.~Santoni, and A.~R. Solomon, ``{Near-zone
  symmetries of Kerr black holes},''
  \href{http://dx.doi.org/10.1007/JHEP09(2022)049}{{\em JHEP} {\bfseries 09}
  (2022) 049}, \href{http://arxiv.org/abs/2203.08832}{{\ttfamily
  arXiv:2203.08832 [hep-th]}}.

\bibitem{Rai:2024lho}
M.~Rai and L.~Santoni, ``{Ladder symmetries and Love numbers of
  Reissner-Nordstr\"om black holes},''
  \href{http://dx.doi.org/10.1007/JHEP07(2024)098}{{\em JHEP} {\bfseries 07}
  (2024) 098}, \href{http://arxiv.org/abs/2404.06544}{{\ttfamily
  arXiv:2404.06544 [gr-qc]}}.

\bibitem{LeTiec:2020spy}
A.~Le~Tiec and M.~Casals, ``{Spinning Black Holes Fall in Love},''
  \href{http://dx.doi.org/10.1103/PhysRevLett.126.131102}{{\em Phys. Rev.
  Lett.} {\bfseries 126} no.~13, (2021) 131102},
  \href{http://arxiv.org/abs/2007.00214}{{\ttfamily arXiv:2007.00214 [gr-qc]}}.

\bibitem{LeTiec:2020bos}
A.~Le~Tiec, M.~Casals, and E.~Franzin, ``{Tidal Love Numbers of Kerr Black
  Holes},'' \href{http://dx.doi.org/10.1103/PhysRevD.103.084021}{{\em Phys.
  Rev. D} {\bfseries 103} no.~8, (2021) 084021},
  \href{http://arxiv.org/abs/2010.15795}{{\ttfamily arXiv:2010.15795 [gr-qc]}}.

\bibitem{Charalambous:2021mea}
P.~Charalambous, S.~Dubovsky, and M.~M. Ivanov, ``{On the Vanishing of Love
  Numbers for Kerr Black Holes},''
  \href{http://dx.doi.org/10.1007/JHEP05(2021)038}{{\em JHEP} {\bfseries 05}
  (2021) 038}, \href{http://arxiv.org/abs/2102.08917}{{\ttfamily
  arXiv:2102.08917 [hep-th]}}.

\bibitem{Rodriguez:2023xjd}
M.~J. Rodriguez, L.~Santoni, A.~R. Solomon, and L.~F. Temoche, ``{Love numbers
  for rotating black holes in higher dimensions},''
  \href{http://dx.doi.org/10.1103/PhysRevD.108.084011}{{\em Phys. Rev. D}
  {\bfseries 108} no.~8, (2023) 084011},
  \href{http://arxiv.org/abs/2304.03743}{{\ttfamily arXiv:2304.03743
  [hep-th]}}.

\bibitem{Goldberger:2006bd}
W.~D. Goldberger and I.~Z. Rothstein, ``{Towers of Gravitational Theories},''
  \href{http://dx.doi.org/10.1142/S0218271806009698}{{\em Gen. Rel. Grav.}
  {\bfseries 38} (2006) 1537--1546},
  \href{http://arxiv.org/abs/hep-th/0605238}{{\ttfamily arXiv:hep-th/0605238}}.

\bibitem{Rothstein:2014sra}
I.~Z. Rothstein, ``{Progress in effective field theory approach to the binary
  inspiral problem},'' \href{http://dx.doi.org/10.1007/s10714-014-1726-y}{{\em
  Gen. Rel. Grav.} {\bfseries 46} (2014) 1726}.

\bibitem{Porto:2016pyg}
R.~A. Porto, ``{The effective field theorist\textquoteright{}s approach to
  gravitational dynamics},''
  \href{http://dx.doi.org/10.1016/j.physrep.2016.04.003}{{\em Phys. Rept.}
  {\bfseries 633} (2016) 1--104},
  \href{http://arxiv.org/abs/1601.04914}{{\ttfamily arXiv:1601.04914
  [hep-th]}}.

\bibitem{Levi:2018nxp}
M.~Levi, ``{Effective Field Theories of Post-Newtonian Gravity: A comprehensive
  review},'' \href{http://dx.doi.org/10.1088/1361-6633/ab12bc}{{\em Rept. Prog.
  Phys.} {\bfseries 83} no.~7, (2020) 075901},
  \href{http://arxiv.org/abs/1807.01699}{{\ttfamily arXiv:1807.01699
  [hep-th]}}.

\bibitem{Goldberger:2022ebt}
W.~D. Goldberger, ``{Effective field theories of gravity and compact binary
  dynamics: A Snowmass 2021 whitepaper},'' in {\em {Snowmass 2021}}.
\newblock 6, 2022.
\newblock \href{http://arxiv.org/abs/2206.14249}{{\ttfamily arXiv:2206.14249
  [hep-th]}}.

\bibitem{Goldberger:2022rqf}
W.~D. Goldberger, ``{Effective Field Theory for Compact Binary Dynamics},''
  \href{http://arxiv.org/abs/2212.06677}{{\ttfamily arXiv:2212.06677
  [hep-th]}}.

\bibitem{Porto:2016zng}
R.~A. Porto, ``{The Tune of Love and the Nature(ness) of Spacetime},''
  \href{http://dx.doi.org/10.1002/prop.201600064}{{\em Fortsch. Phys.}
  {\bfseries 64} no.~10, (2016) 723--729},
  \href{http://arxiv.org/abs/1606.08895}{{\ttfamily arXiv:1606.08895 [gr-qc]}}.

\bibitem{BenAchour:2022uqo}
J.~Ben~Achour, E.~R. Livine, S.~Mukohyama, and J.-P. Uzan, ``{Hidden symmetry
  of the static response of black holes: applications to Love numbers},''
  \href{http://dx.doi.org/10.1007/JHEP07(2022)112}{{\em JHEP} {\bfseries 07}
  (2022) 112}, \href{http://arxiv.org/abs/2202.12828}{{\ttfamily
  arXiv:2202.12828 [gr-qc]}}.

\bibitem{Berens:2022ebl}
R.~Berens, L.~Hui, and Z.~Sun, ``{Ladder symmetries of black holes and de
  Sitter space: love numbers and quasinormal modes},''
  \href{http://dx.doi.org/10.1088/1475-7516/2023/06/056}{{\em JCAP} {\bfseries
  06} (2023) 056}, \href{http://arxiv.org/abs/2212.09367}{{\ttfamily
  arXiv:2212.09367 [hep-th]}}.

\bibitem{Charalambous:2021kcz}
P.~Charalambous, S.~Dubovsky, and M.~M. Ivanov, ``{Hidden Symmetry of Vanishing
  Love Numbers},'' \href{http://dx.doi.org/10.1103/PhysRevLett.127.101101}{{\em
  Phys. Rev. Lett.} {\bfseries 127} no.~10, (2021) 101101},
  \href{http://arxiv.org/abs/2103.01234}{{\ttfamily arXiv:2103.01234
  [hep-th]}}.

\bibitem{Charalambous:2022rre}
P.~Charalambous, S.~Dubovsky, and M.~M. Ivanov, ``{Love symmetry},''
  \href{http://dx.doi.org/10.1007/JHEP10(2022)175}{{\em JHEP} {\bfseries 10}
  (2022) 175}, \href{http://arxiv.org/abs/2209.02091}{{\ttfamily
  arXiv:2209.02091 [hep-th]}}.

\bibitem{Solomon:2023ltn}
A.~R. Solomon, ``{Off-Shell Duality Invariance of Schwarzschild Perturbation
  Theory},'' \href{http://dx.doi.org/10.3390/particles6040061}{{\em Particles}
  {\bfseries 6} no.~4, (2023) 943--974},
  \href{http://arxiv.org/abs/2310.04502}{{\ttfamily arXiv:2310.04502 [gr-qc]}}.

\bibitem{Sharma:2024hlz}
C.~Sharma, R.~Ghosh, and S.~Sarkar, ``{Exploring ladder symmetry and Love
  numbers for static and rotating black holes},''
  \href{http://dx.doi.org/10.1103/PhysRevD.109.L041505}{{\em Phys. Rev. D}
  {\bfseries 109} no.~4, (2024) L041505},
  \href{http://arxiv.org/abs/2401.00703}{{\ttfamily arXiv:2401.00703 [gr-qc]}}.

\bibitem{Poisson:2020vap}
E.~Poisson, ``{Compact body in a tidal environment: New types of relativistic
  Love numbers, and a post-Newtonian operational definition for tidally induced
  multipole moments},''
  \href{http://dx.doi.org/10.1103/PhysRevD.103.064023}{{\em Phys. Rev. D}
  {\bfseries 103} no.~6, (2021) 064023},
  \href{http://arxiv.org/abs/2012.10184}{{\ttfamily arXiv:2012.10184 [gr-qc]}}.

\bibitem{Poisson:2021yau}
E.~Poisson, ``{Tidally induced multipole moments of a nonrotating black hole
  vanish to all post-Newtonian orders},''
  \href{http://dx.doi.org/10.1103/PhysRevD.104.104062}{{\em Phys. Rev. D}
  {\bfseries 104} no.~10, (2021) 104062},
  \href{http://arxiv.org/abs/2108.07328}{{\ttfamily arXiv:2108.07328 [gr-qc]}}.

\bibitem{Riva:2023rcm}
M.~M. Riva, L.~Santoni, N.~Savi\'c, and F.~Vernizzi, ``{Vanishing of nonlinear
  tidal Love numbers of Schwarzschild black holes},''
  \href{http://dx.doi.org/10.1016/j.physletb.2024.138710}{{\em Phys. Lett. B}
  {\bfseries 854} (2024) 138710},
  \href{http://arxiv.org/abs/2312.05065}{{\ttfamily arXiv:2312.05065 [gr-qc]}}.

\bibitem{Iteanu:2024dvx}
S.~Iteanu, M.~M. Riva, L.~Santoni, N.~Savi\'c, and F.~Vernizzi, ``{Vanishing of
  Quadratic Love Numbers of Schwarzschild Black Holes},''
  \href{http://arxiv.org/abs/2410.03542}{{\ttfamily arXiv:2410.03542 [gr-qc]}}.

\bibitem{DeLuca:2023mio}
V.~De~Luca, J.~Khoury, and S.~S.~C. Wong, ``{Nonlinearities in the tidal Love
  numbers of black holes},''
  \href{http://dx.doi.org/10.1103/PhysRevD.108.024048}{{\em Phys. Rev. D}
  {\bfseries 108} no.~2, (2023) 024048},
  \href{http://arxiv.org/abs/2305.14444}{{\ttfamily arXiv:2305.14444 [gr-qc]}}.

\bibitem{Perry:2023wmm}
M.~Perry and M.~J. Rodriguez, ``{Dynamical Love Numbers for Kerr Black
  Holes},'' \href{http://arxiv.org/abs/2310.03660}{{\ttfamily arXiv:2310.03660
  [gr-qc]}}.

\bibitem{Griffiths:2009dfa}
J.~B. Griffiths and J.~Podolsky,
  \href{http://dx.doi.org/10.1017/CBO9780511635397}{{\em {Exact Space-Times in
  Einstein's General Relativity}}}.
\newblock Cambridge Monographs on Mathematical Physics. Cambridge University
  Press, Cambridge, 2009.

\bibitem{Townsend:1997ku}
P.~K. Townsend, ``{Black holes: Lecture notes},''
  \href{http://arxiv.org/abs/gr-qc/9707012}{{\ttfamily arXiv:gr-qc/9707012}}.

\bibitem{Weyl:1917gp}
H.~Weyl, ``{The theory of gravitation},''
  \href{http://dx.doi.org/10.1007/s10714-011-1310-7}{{\em Annalen Phys.}
  {\bfseries 54} (1917) 117--145}.

\bibitem{Stephani:2003tm}
H.~Stephani, D.~Kramer, M.~A.~H. MacCallum, C.~Hoenselaers, and E.~Herlt,
  \href{http://dx.doi.org/10.1017/CBO9780511535185}{{\em {Exact solutions of
  Einstein's field equations}}}.
\newblock Cambridge Monographs on Mathematical Physics. Cambridge Univ. Press,
  Cambridge, 2003.

\bibitem{Barcelo:2024ioe}
C.~Barcel\'o, R.~Carballo-Rubio, L.~J. Garay, and G.~Garc\'\i{}a-Moreno,
  ``{No-hair and almost-no-hair results for static axisymmetric black holes and
  ultracompact objects in astrophysical environments},''
  \href{http://arxiv.org/abs/2410.08128}{{\ttfamily arXiv:2410.08128 [gr-qc]}}.

\bibitem{Gurlebeck:2013eia}
N.~G\"urlebeck, ``{Source integrals of asymptotic multipole moments},''
  \href{http://dx.doi.org/10.1007/978-3-319-06761-2_11}{{\em Springer Proc.
  Phys.} {\bfseries 157} (2014) 83--90},
  \href{http://arxiv.org/abs/1302.7234}{{\ttfamily arXiv:1302.7234 [gr-qc]}}.

\bibitem{Geroch:1970cc}
R.~P. Geroch, ``{Multipole moments. I. Flat space},''
  \href{http://dx.doi.org/10.1063/1.1665348}{{\em J. Math. Phys.} {\bfseries
  11} (1970) 1955--1961}.

\bibitem{Geroch:1970cd}
R.~P. Geroch, ``{Multipole moments. II. Curved space},''
  \href{http://dx.doi.org/10.1063/1.1665427}{{\em J. Math. Phys.} {\bfseries
  11} (1970) 2580--2588}.

\bibitem{Hansen:1974zz}
R.~O. Hansen, ``{Multipole moments of stationary space-times},''
  \href{http://dx.doi.org/10.1063/1.1666501}{{\em J. Math. Phys.} {\bfseries
  15} (1974) 46--52}.

\bibitem{Mayerson:2022ekj}
D.~R. Mayerson, ``{Gravitational multipoles in general stationary
  spacetimes},'' \href{http://dx.doi.org/10.21468/SciPostPhys.15.4.154}{{\em
  SciPost Phys.} {\bfseries 15} no.~4, (2023) 154},
  \href{http://arxiv.org/abs/2210.05687}{{\ttfamily arXiv:2210.05687 [gr-qc]}}.

\bibitem{Bern:2020uwk}
Z.~Bern, J.~Parra-Martinez, R.~Roiban, E.~Sawyer, and C.-H. Shen, ``{Leading
  Nonlinear Tidal Effects and Scattering Amplitudes},''
  \href{http://dx.doi.org/10.1007/JHEP05(2021)188}{{\em JHEP} {\bfseries 05}
  (2021) 188}, \href{http://arxiv.org/abs/2010.08559}{{\ttfamily
  arXiv:2010.08559 [hep-th]}}.

\bibitem{Ehlers:1957zz}
J.~Ehlers, ``{Konstruktionen und Charakterisierung von Losungen der
  Einsteinschen Gravitationsfeldgleichungen},'' other thesis, 1957.

\bibitem{Geroch:1970nt}
R.~P. Geroch, ``{A Method for generating solutions of Einstein's equations},''
  \href{http://dx.doi.org/10.1063/1.1665681}{{\em J. Math. Phys.} {\bfseries
  12} (1971) 918--924}.

\bibitem{Breitenlohner:1986um}
P.~Breitenlohner and D.~Maison, ``{On the Geroch Group},'' {\em Ann. Inst. H.
  Poincare Phys. Theor.} {\bfseries 46} (1987) 215.

\bibitem{Maison:2000fj}
D.~Maison, ``{Duality and hidden symmetries in gravitational theories},'' {\em
  Lect. Notes Phys.} {\bfseries 540} (2000) 273--323.

\bibitem{Lu:2007zv}
H.~Lu, M.~J. Perry, and C.~N. Pope, ``{Infinite-dimensional symmetries of
  two-dimensional coset models},''
  \href{http://arxiv.org/abs/0711.0400}{{\ttfamily arXiv:0711.0400 [hep-th]}}.

\bibitem{Lu:2007jc}
H.~Lu, M.~J. Perry, and C.~N. Pope, ``{Infinite-dimensional symmetries of
  two-dimensional coset models coupled to gravity},''
  \href{http://dx.doi.org/10.1016/j.nuclphysb.2008.07.035}{{\em Nucl. Phys. B}
  {\bfseries 806} (2009) 656--683},
  \href{http://arxiv.org/abs/0712.0615}{{\ttfamily arXiv:0712.0615 [hep-th]}}.

\bibitem{Maison:1978es}
D.~Maison, ``{Are the stationary, axially symmetric Einstein equations
  completely integrable?},''
  \href{http://dx.doi.org/10.1103/PhysRevLett.41.521}{{\em Phys. Rev. Lett.}
  {\bfseries 41} (1978) 521}.

\bibitem{Schwarz:1995td}
J.~H. Schwarz, ``{Classical symmetries of some two-dimensional models},''
  \href{http://dx.doi.org/10.1016/0550-3213(95)00276-X}{{\em Nucl. Phys. B}
  {\bfseries 447} (1995) 137--182},
  \href{http://arxiv.org/abs/hep-th/9503078}{{\ttfamily arXiv:hep-th/9503078}}.

\bibitem{Schwarz:1995af}
J.~H. Schwarz, ``{Classical symmetries of some two-dimensional models coupled
  to gravity},'' \href{http://dx.doi.org/10.1016/0550-3213(95)00455-2}{{\em
  Nucl. Phys. B} {\bfseries 454} (1995) 427--448},
  \href{http://arxiv.org/abs/hep-th/9506076}{{\ttfamily arXiv:hep-th/9506076}}.

\bibitem{Kehagias:2024rtz}
A.~Kehagias and A.~Riotto, ``{Black Holes in a Gravitational Field: The
  Non-linear Static Love Number of Schwarzschild Black Holes Vanishes},''
  \href{http://arxiv.org/abs/2410.11014}{{\ttfamily arXiv:2410.11014 [gr-qc]}}.

\bibitem{Geroch:1982bv}
R.~P. Geroch and J.~B. Hartle, ``{Distorted black holes},''
  \href{http://dx.doi.org/10.1063/1.525384}{{\em J. Math. Phys.} {\bfseries 23}
  (1982) 680}.

\bibitem{Miller_1984}
W.~Miller, {\em The Three-Variable Helmholtz and Laplace Equations},
  p.~160–222.
\newblock Encyclopedia of Mathematics and its Applications.
\newblock Cambridge University Press, 1984.

\bibitem{Regge:1957td}
T.~Regge and J.~A. Wheeler, ``{Stability of a Schwarzschild singularity},''
  \href{http://dx.doi.org/10.1103/PhysRev.108.1063}{{\em Phys. Rev.} {\bfseries
  108} (1957) 1063--1069}.

\bibitem{Foffa:2013qca}
S.~Foffa and R.~Sturani, ``{Effective field theory methods to model compact
  binaries},'' \href{http://dx.doi.org/10.1088/0264-9381/31/4/043001}{{\em
  Class. Quant. Grav.} {\bfseries 31} no.~4, (2014) 043001},
  \href{http://arxiv.org/abs/1309.3474}{{\ttfamily arXiv:1309.3474 [gr-qc]}}.

\bibitem{Goldberger:2020fot}
W.~D. Goldberger, J.~Li, and I.~Z. Rothstein, ``{Non-conservative effects on
  spinning black holes from world-line effective field theory},''
  \href{http://dx.doi.org/10.1007/JHEP06(2021)053}{{\em JHEP} {\bfseries 06}
  (2021) 053}, \href{http://arxiv.org/abs/2012.14869}{{\ttfamily
  arXiv:2012.14869 [hep-th]}}.

\bibitem{Ivanov:2022hlo}
M.~M. Ivanov and Z.~Zhou, ``{Revisiting the matching of black hole tidal
  responses: A systematic study of relativistic and logarithmic corrections},''
  \href{http://dx.doi.org/10.1103/PhysRevD.107.084030}{{\em Phys. Rev. D}
  {\bfseries 107} no.~8, (2023) 084030},
  \href{http://arxiv.org/abs/2208.08459}{{\ttfamily arXiv:2208.08459
  [hep-th]}}.

\bibitem{Hadad:2024lsf}
T.~Hadad, B.~Kol, and M.~Smolkin, ``{Gravito-magnetic polarization of
  Schwarzschild black hole},''
  \href{http://dx.doi.org/10.1007/JHEP06(2024)169}{{\em JHEP} {\bfseries 06}
  (2024) 169}, \href{http://arxiv.org/abs/2402.16172}{{\ttfamily
  arXiv:2402.16172 [hep-th]}}.

\bibitem{10.1093/acprof:oso/9780198702764.001.0001}
G.~S. He,
  \href{http://dx.doi.org/10.1093/acprof:oso/9780198702764.001.0001}{{\em
  {Nonlinear Optics and Photonics}}}.
\newblock Oxford University Press, 10, 2014.
\newblock \url{https://doi.org/10.1093/acprof:oso/9780198702764.001.0001}.

\bibitem{Saketh:2023bul}
M.~V.~S. Saketh, Z.~Zhou, and M.~M. Ivanov, ``{Dynamical tidal response of Kerr
  black holes from scattering amplitudes},''
  \href{http://dx.doi.org/10.1103/PhysRevD.109.064058}{{\em Phys. Rev. D}
  {\bfseries 109} no.~6, (2024) 064058},
  \href{http://arxiv.org/abs/2307.10391}{{\ttfamily arXiv:2307.10391
  [hep-th]}}.

\bibitem{Haddad:2020que}
K.~Haddad and A.~Helset, ``{Tidal effects in quantum field theory},''
  \href{http://dx.doi.org/10.1007/JHEP12(2020)024}{{\em JHEP} {\bfseries 12}
  (2020) 024}, \href{http://arxiv.org/abs/2008.04920}{{\ttfamily
  arXiv:2008.04920 [hep-th]}}.

\bibitem{Ruhdorfer:2019qmk}
M.~Ruhdorfer, J.~Serra, and A.~Weiler, ``{Effective Field Theory of Gravity to
  All Orders},'' \href{http://dx.doi.org/10.1007/JHEP05(2020)083}{{\em JHEP}
  {\bfseries 05} (2020) 083}, \href{http://arxiv.org/abs/1908.08050}{{\ttfamily
  arXiv:1908.08050 [hep-ph]}}.

\bibitem{Poisson_Will_2014}
E.~Poisson and C.~M. Will,
  \href{http://dx.doi.org/10.1017/CBO9781139507486}{{\em Gravity: Newtonian,
  Post-Newtonian, Relativistic}}.
\newblock Cambridge University Press, 2014.

\bibitem{Compton:2020cjx}
G.~Compton and I.~A. Morrison, ``{Hidden symmetries for transparent de Sitter
  space},'' \href{http://dx.doi.org/10.1088/1361-6382/ab8c98}{{\em Class.
  Quant. Grav.} {\bfseries 37} no.~12, (2020) 125001},
  \href{http://arxiv.org/abs/2003.08023}{{\ttfamily arXiv:2003.08023 [gr-qc]}}.

\bibitem{Katsimpouri:2012ky}
D.~Katsimpouri, A.~Kleinschmidt, and A.~Virmani, ``{Inverse Scattering and the
  Geroch Group},'' \href{http://dx.doi.org/10.1007/JHEP02(2013)011}{{\em JHEP}
  {\bfseries 02} (2013) 011}, \href{http://arxiv.org/abs/1211.3044}{{\ttfamily
  arXiv:1211.3044 [hep-th]}}.

\bibitem{Katsimpouri:2015nqc}
D.~Katsimpouri, {\em {Integrability in two-dimensional gravity}}.
\newblock PhD thesis, Humboldt U., Berlin, 2015.

\bibitem{Cardoso:2017cgi}
G.~L. Cardoso and J.~C. Serra, ``{New gravitational solutions via a
  Riemann-Hilbert approach},''
  \href{http://dx.doi.org/10.1007/JHEP03(2018)080}{{\em JHEP} {\bfseries 03}
  (2018) 080}, \href{http://arxiv.org/abs/1711.01113}{{\ttfamily
  arXiv:1711.01113 [hep-th]}}.

\bibitem{Poisson:2009qj}
E.~Poisson and I.~Vlasov, ``{Geometry and dynamics of a tidally deformed black
  hole},'' \href{http://dx.doi.org/10.1103/PhysRevD.81.024029}{{\em Phys. Rev.
  D} {\bfseries 81} (2010) 024029},
  \href{http://arxiv.org/abs/0910.4311}{{\ttfamily arXiv:0910.4311 [gr-qc]}}.

\bibitem{Hinderer:2007mb}
T.~Hinderer, ``{Tidal Love numbers of neutron stars},''
  \href{http://dx.doi.org/10.1086/533487}{{\em Astrophys. J.} {\bfseries 677}
  (2008) 1216--1220}, \href{http://arxiv.org/abs/0711.2420}{{\ttfamily
  arXiv:0711.2420 [astro-ph]}}.

\bibitem{Gurlebeck:2012mb}
N.~Gurlebeck, ``{Source integrals for multipole moments in static and axially
  symmetric spacetimes},''
  \href{http://dx.doi.org/10.1103/PhysRevD.90.024041}{{\em Phys. Rev. D}
  {\bfseries 90} no.~2, (2014) 024041},
  \href{http://arxiv.org/abs/1207.4500}{{\ttfamily arXiv:1207.4500 [gr-qc]}}.

\end{thebibliography}\endgroup
}

\end{document}